\documentclass[apj]{emulateapj}

\def\Msun{$M_\odot$}
\def\msun{$M_\odot$}

\newcommand{\chem}[2]{$\mathrm{^{#1}#2}$}
\usepackage{rotating}

\begin{document} 

\title{Radiation-hydrodynamical modelling of Core-Collapse Supernovae:
light curves and the evolution of photospheric velocity and temperature}

\author{M.~L. Pumo\altaffilmark{1,2,3} \& L. Zampieri\altaffilmark{1}} 
\shortauthors{Pumo \& Zampieri}
\affil{\altaffilmark{1}INAF-Osservatorio Astronomico di Padova, Vicolo dell'Osservatorio 5, I-35122 Padova, Italy}
\affil{\altaffilmark{2}Bonino-Pulejo Foundation, Via Uberto Bonino 15/C, I-98124 Messina, Italy}
\affil{\altaffilmark{3}INAF-Osservatorio Astrofisico di Catania, via S. Sofia 78, I-95123 Catania, Italy}
\date{Received ... ; accepted ...}

\begin{abstract}
We have developed a relativistic, radiation-hydrodynamics Lagrangian code, specifically tailored 
to simulate the evolution of the main observables (light curve, evolution of photospheric velocity 
and temperature) in core-collapse supernova (CC-SN) events. The distinctive features of the code
are an accurate treatment of radiative transfer coupled to relativistic hydrodynamics, a self-consistent 
treatment of the evolution of the innermost ejecta taking into account the gravitational effects of the 
central compact remnant, and a fully implicit Lagrangian approach to the solution of the coupled 
non-linear finite difference system of equations. Our aim is to use it as numerical tool 
to perform calculations of grid of models to be compared with observation of CC-SNe. 
In this paper we present some testcase simulations and a comparison with observations of SN 1987A, 
as well as with the results obtained with other numerical codes. We also briefly discuss the influence 
of the main physical parameters (ejected mass, progenitor radius, explosion energy, amount of \chem{56}{Ni}) 
on the evolution of the ejecta, and the implications of our results in connection with the possibility to 
``standardize'' hydrogen-rich CC-SNe for using them as candles to measure cosmological distances.
\end{abstract}

\keywords{supernovae: general --- hydrodynamics --- radiative transfer --- methods: numerical --- 
supernovae: individual (SN 1987A) --- distance scale}

\section{Introduction}
\label{sec:intro}

It is widely accepted that core-collapse supernovae (CC-SNe) represent the final explosive
evolutionary phase of stars having initial (i.e. main sequence) masses larger than $\sim$ 
8-10\Msun\, \citep[e.g.][]{woosley86,hamuy03,heger03}. As such, CC-SNe are fundamental 
probes of the stellar evolution of massive stars and can be used to understand the link 
among explosion mechanism, nature of remnants, progenitors evolution, and environment 
around progenitors \citep[e.g.][]{filip97,CT00}.

In addition to their intrinsic interest, CC-SNe are relevant to many astrophysical issues. 
For example, the CC-SNe influence the physical evolution of galaxies, their frequency
being related to the on-going star formation rate \citep*[e.g.][]{cappellaro05,madau98} and
determining the kinematics and the structure of the interstellar medium \citep[e.g.][]{RB89}. 
CC-SNe are also important because of their role in the production of neutrinos, cosmic rays
and, probably, gravitational waves \citep*[e.g.][]{haungs03,DK09,pagliaroli09}. Additionally, 
the CC-SN ejecta, enriched in heavy elements, make CC-SNe key actors in the nucleosynthesis 
processes of intermediate and trans-iron elements and in the chemical evolution of galaxies 
\citep*[e.g.][]{arnett96,chieffi98}. Moreover, they seem to be particularly promising to 
measure cosmological distances in addition to type Ia SNe 
\citep[e.g.][and references therein]{nugent06,zampieri07,olivares10}.

In spite of the importance of these explosive events in astrophysics, there are still basic 
questions to be answered on CC-SNe, related to the extreme variety of their displays. Indeed 
they appear to show different energetics and to eject rather diverse amounts of material, 
causing rather heterogeneous behaviors of their light curves and spectra \citep*[e.g.][]{turatto07}. 
This apparent heterogeneity may be linked to stellar evolution effects and to the explosion mechanism 
\citep[e.g.][and references therein]{pumo09b}, but the exact link between the physical 
properties of the explosion (ejected mass, explosion energy, stellar structure and composition 
at the explosion) and the observational characteristics is far from being well-established. 

Light curve and spectral modelling have often successfully been used to provide information 
about the physical properties of single CC-SNe \citep*[e.g. SNe 1987A, 1993J, 1999em, 2004et, 
and 2005cs; see][]{shige88,woosley88,shigenomoto90,blinnikov98,utrobin04,utrobin07,baklanov05,
UC08,pasto09,bersten11}. Comparatively little effort has been devoted to investigating 
large samples of these events and trying to explain the significant range of properties that they 
show. Differences in luminosity, expansion velocity and \chem{56}{Ni} yields are very large (up to 
$\sim$ 2 order of magnitude), and difficult to relate to changes in a single parameter, 
such as the mass of the progenitor star. However, despite this, variations in the observed properties 
appear to obey a certain order. Some correlations among different observables have been noted 
\citep[e.g.][]{hamuy03b,zampieri07} and used to calibrate CC-SNe as distance indicators 
\citep[e.g.][]{HP02,hamuy03b,nugent06,olivares10}, in addition to Baade-Wesselink type, spectral 
based techniques (like the expanding photosphere method, see e.g. \citealt{KK74}; 
\citealt*{eastman96}; \citealt{dessart08}; or the spectral-fitting expanding atmosphere method, see 
e.g. \citealt{mitchell02}; \citealt{baron04}).

While parameter studies of the light curves of CC-SNe have been largely pursued in
the past both with analytical \citep[e.g.][and references therein]{BP93,popov93,arnett96} 
and numerical \citep*[e.g.][]{LN83,LN85,Y04,dessart10} techniques, only a few 
systematic theoretical investigations of the physical origin of the aforementioned 
correlations have been performed \citep[e.g.][]{zampieri07,KW09}. It is still not 
completely clear how variations of basic parameters conspire in such a way to give rise 
to the observed relations. In particular, \citet{KW09} found that the relation between 
luminosity and expansion velocity at 50 days is explained by the behavior of hydrogen 
recombination in the SN envelope and proposed additional correlations based only on 
photometric data. A better physical understanding of these correlations would allow us, 
on one side, to pinpoint possible biases that may affect the use of CC-SNe as standard 
candles and, on the other side, to look for other correlations that may reveal more 
promising when applied to high redshift SNe. 

With the aim of clarifying the aforementioned astrophysical issues, we have improved on previous 
work that we started on the numerical modelling of SN ejecta and fallback \citep*{zampieri98,balberg00}, 
and on the physical interpretation of the heterogeneous behavior of Type II plateau SNe 
\citep{zampieri03,zampieri05,zampieri07}. In this paper we present a new, improved version of a 
general-relativistic, radiation hydrodynamics, Lagrangian code tailored to the modelling of CC-SNe. 
Similar numerical treatment of the expanding SN envelope have been presented in the literature 
\citep[e.g.][]{blinnikov98,iwamoto00,chieffi03,Y04,KW09,bersten11}. Our code has the advantage of being 
able to follow the fallback of material on the central remnant in a fully relativistic formalism, and 
allows for an accurate treatment of radiative transfer in all regimes. It will enable us to explore the 
physical behavior of CC-SNe, to link their physical and observational properties, to investigate the 
existence of correlations among their different observables, and also to perform model fitting of single events.

The plan of the paper is the following. In sect.~\ref{sec:code} we describe the features of 
the new version of the code (details on the adopted numerical procedure and on the finite 
difference form of the equations are presented in the Appendix). Sect.~\ref{sec:simul} 
illustrates the numerical calculations. Sect.~\ref{sec:validation} is devoted to the code validation 
and the analysis of the basic physical properties of the ejecta in presence of a central compact 
remnant during all the different evolutionary stages. A summary with final comments and a short 
discussion of the implications of our results in connection with the possibility to ``standardizate'' 
hydrogen-rich CC-SNe is presented in Sect.~\ref{sec:summary}.


\section{Code description}
\label{sec:code}

The code is a new, improved version of a general-relativistic, radiation hydrodynamics, 
Lagrangian code described in \citet*[][]{zampieri96}, \citet{zampieri98} and in \citet[][]{balberg00}, 
and originally developed for studying spherical accretion onto black holes and fall back in 
the aftermath of a SN explosion.

This new version of the code is able to compute the evolution of the ejecta and the emitted 
luminosity, from the breakout of the shock wave at the stellar surface up to the nebular stage 
(when the envelope has recombined and the energy budget is dominated by the radioactive decays 
of the heavy elements synthesized in the explosion). We solve the equations of relativistic 
radiation hydrodynamics in spherically symmetry for a self-gravitating matter fluid which interacts 
with radiation, taking into account the heating effects due to decays of radioactive isotopes 
synthesized in the SN explosion. The distinctive features of the code are summarized here, while a 
detailed description of the equations, numerical method and input physics is reported below (see 
sections \ref{sec:code:eq} through \ref{sec:code:physics}). The code encompasses:
\begin{itemize}
 \item[-] an accurate treatment of radiative transfer in all regimes from the one in which the ejecta 
are optically thick up to when they are completely transparent. This is obtained through the 
solution of the first two moments of the radiative transfer equations (Eqs.~[\ref{eq:w0}] and [\ref{eq:w1}] 
in sect.~\ref{sec:code:eq}). Although it is necessary to adopt a closure relation, which is constrained
only in the asymptotic optically thick/thin regimes \citep[see][for more details]{zampieri96,zampieri98},
the uncertainty induced by this approximation is not significant for wavelength(frequency)-integrated radiative 
transfer \citep[see e.g.][]{NT88} and this approach is definitely superior to any treatment based on the 
diffusion approximation;
 \item[-] the coupling of the radiation moment equations with the equations of relativistic hydrodynamics, 
adopting a fully implicit Lagrangian finite difference scheme (see sect.~\ref{sec:code:eq} and \ref{sec:code:num} 
for details);
 \item[-] the possibility to compute the evolution of the ejecta and the emitted luminosity taking into account 
the gravitational effects of the compact remnant. This enables the code to deal with the fallback of material 
onto the compact remnant and, consequently, to accurately determining the amount of ejected \chem{56}{Ni}.
\end{itemize}

The main limitation of our code compared to other similar codes (e.g. \citealt{Y04}, \citealt{KW09}, and 
\citealt*{dessart10}) is that we consider only wavelength integrated radiation energy density and luminosity, 
and hence we cannot compute the actual spectrum of the SN. Also, the evolution starts from ``ad hoc'' initial 
conditions that mimic the profiles of the physical variables after shock and reverse shock passage, but such 
conditions are not meant to be ``perfectly realistic'', in the sense to be the outcome of a stellar evolution 
code coupled with numerical computations of the explosion phase. Some improvements in this respect have been 
already added to the present version of the code (see sect.~\ref{sec:code:IniCond} for details). At the same 
time, we are working (in collaboration with A.~Chieffi and M.~Limongi) at interfacing it with the output of a 
hydrodynamical code that follows the CC-SN explosion and the explosive nucleosynthesis, starting from pre-SN 
models obtained through a stellar evolutionary code.

\subsection{Equations}
\label{sec:code:eq}

Since we consider the effects of the central compact remnant on the evolution of the ejecta, we adopt a 
fully general relativistic treatment. The equations governing the relativistic radiation hydrodynamics of 
the expanding ejecta in spherically symmetry are reported in \citet{zampieri96,zampieri98} 
and \citet{balberg00}. The main physical variables that describe the behavior of the ejecta are the gas 
mass density $\rho$, and the gas internal energy per unit mass $\epsilon$ and pressure $p$ (measured in the frame 
comoving with the ejecta). Those describing the radiation field are the radiation energy density $w_0$ and 
flux $w_1$ (both in units of erg cm$^{-3}$). We refer to \citet{zampieri98} for the full set of equations and 
related quantities, while here we summarize for convenience the gas energy equation and the first two moments 
of the radiative transfer equations, whose treatment has been deeply modified in the present analysis. They read:
\begin{equation}
 \label{eq:energy}
 \epsilon_{,t} + ak_P(B-w_0) + p\left(\frac{1}{\rho}\right)_{,t} = Q \,\,\,\,\mbox{energy eq.,}
\end{equation}

\begin{eqnarray}
 \label{eq:w0}
 (w_0)_{,t} - ak_P\rho(B-w_0) + \nonumber \\ 
 w_0 \left[\frac{4}{3} \left( \frac{b_{,t}}{b} + 2 \frac{r_{,t}}{r} \right) + f \left(\frac{b_{,t}}{b}-\frac{r_{,t}}{r}\right) \right] + \nonumber \\ 
 \frac{1}{ab{r^2}}(w_1 a^2 r^2)_{,\mu}= 0 \,\,\,\,\mbox{zero-th moment eq.,} 
\end{eqnarray}

\begin{eqnarray}
 \label{eq:w1}
 (w_1)_{,t} + ak_R\rho w_1 + \nonumber \\ 
 2w_1\left( \frac{b_{,t}}{b} + \frac{r_{,t}}{r} \right) + a\left[\frac{1}{3a^4b}(w_0a^4)_{,\mu} +  \right. \nonumber \\
 \left. \frac{1}{3br^3}(fw_0ar^3)_{,\mu} \right]= 0 \,\,\,\,\mbox{first moment eq.,} 
\end{eqnarray}
where $t$ and $\mu$ are the Lagrangian time and mass contained within a comoving spherical shell of 
radius $r$ (a comma denotes partial derivates with respect to the corresponding variable), $a$ (00 metric 
coefficient) is a function of $t$ and $\mu$ (computed from Eq.~[6] in \citealt{zampieri98}), 
$b=1/(4\pi r^2 \rho)$, $B=a_R T^4$ ($a_R$ blackbody radiation constant), $Q$ is the heating rate of 
radioactive decays from the isotopes synthesized in the SN explosion (see sect.~\ref{sec:code:physics}), 
and $k_P$ and $k_R$ are Planck mean and Rosseland mean opacity, respectively. The dependence on the gas 
temperature $T$ is contained in $\epsilon$, $p$, $k_P$ and $k_R$, and is specified through the equations 
of state (see sect.~\ref{sec:code:physics}). $k_P$ and $k_R$ are calculated interpolating the TOPS opacities 
as a function of $T$ and $\rho$ (see again sect.~\ref{sec:code:physics}). Finally, the function $f$ (Eddington 
factor), relating the second-order moment of the radiation intensity to $w_0$, is calculated using Eqs.~[12] 
and [13] in \citet{zampieri98}.

\subsection{Numerical method}
\label{sec:code:num}

In order to solve the complete system of equations of relativistic radiation hydrodynamics, the old version 
of the code adopts a semi-implicit Lagrangian finite difference scheme where the time step is controlled by 
the Courant condition and the requirement that the fractional variation of the variables in one time-step 
be smaller than 10\%. The energy equation (Eq.~[\ref{eq:energy}]) and the zero-th moment of the radiative 
transfer equation (Eq.~[\ref{eq:w0}]) form a non-linear system of equations in the gas temperature $T$ 
and are then solved point-by-point on the computational grid using a Newton-Raphson iterative method 
\citep[see the Appendix in][for details]{zampieri98}.

We have deeply modified the numerical treatment previously adopted by implementing a fully implicit Lagrangian 
finite difference scheme. This allows for a major improvement in the numerical stability and overall computational 
efficiency of the code, especially during those phases when fast motions of steep gradients occur (e.g. the 
radiative recombination phase). The first moment equation for the radiative flux $w_1$ (Eq.~[\ref{eq:w1}]) is now 
solved together with Eqs.~[\ref{eq:energy}] and [\ref{eq:w0}] in a fully implicit scheme on the whole computational 
grid, that requires a modification of the temporal centering of the variables. This leads to a highly non-linear 
system of equations that is solved through a Newton-Raphson iterative method and matrix inversion packages that 
minimise the CPU time and the required storage space. In the Appendix we report more details on the finite 
difference form of the Eqs.~[\ref{eq:energy}], [\ref{eq:w0}] and [\ref{eq:w1}], and the numerical procedure 
adopted to solve them.

\subsection{Initial and boundary conditions}
\label{sec:code:IniCond}

In the present analysis we consider the evolution of the ejecta after the explosion phase{\footnote{We are 
working at interfacing our code with the output of self-consistent calculations of the explosion phase, but 
during this testing phase we preferred to assume idealized initial conditions that provide an approximate 
description of the ejected material after shock (and possible reverse shock) passage. Clearly, as the actual 
velocity, density and heavy element distributions of the post-shock material may affect the light curve and 
the determination of the envelope's physical parameters in a significant way, at present we can provide only 
approximate estimates of these quantities.}. At this stage, the SN shock wave has already propagated through the envelope 
of the progenitor star redistributing the explosion energy through it, and the evolution starts when the 
envelope is essentially free-coasting and in homologous expansion. We assume that it is comprised of a 
mixture of hydrogen, helium and heavier elements. As oxygen is expected to be the most abundant heavy element, 
we further assume that it represents the entire metal component in the envelope's final composition 
\citep[see][]{balberg00}. The mass fraction of \chem{56}{Ni} is assumed to be concentrated 
towards the center and to vary as a function of Lagrangian mass $\mu$ as:
\begin{equation}
\label{eq:Ni56}
X_{\mathrm{^{56}{Ni}}}(\mu)=X_{\mathrm{^{56}{Ni,0}}}\times exp{\left[-K_{mix}\frac{\mu}{\mu_{max}}\right]} \,\mbox{,}
\end{equation}
where $X_{\mathrm{^{56}{Ni,0}}}$ is the initial mass fraction of \chem{56}{Ni} at the inner boundary 
of the ejecta, $K_{mix}$ is a numerical constant whose value is essentially related to the extent of mixing 
of \chem{56}{Ni} throughout the envelope, and $\mu_{max}$ is the value of $\mu$ at the outer boundary, equal 
to the total envelope mass (see Appendix \ref{appendix} for details).

Assuming that, at the onset of expansion, radiation is in local thermodynamic equilibrium (LTE hereafter) with 
the gas throughout the envelope and the initial photon mean free path $\lambda_0$ is much smaller than the
radius of the envelope, the initial temperature profile in the ejecta can be well approximated by the 
so-called ``radiative zero solution'' of \citet{arnett80}:
\begin{equation}
\label{eq:Temp}
T(r(\mu),t=0) = T_0 \times \left[ \frac{sin(\pi x)}{\pi x}\right]^{\frac{1}{4}}\,\mbox{,}
\end{equation}
where $T_0$ is the initial central temperature, $x = r/R_0$ and $R_0 \equiv r(\mu_{max})$ is the initial radius of 
the outermost shell of the ejecta. 

As the gas is initially in LTE with radiation, we assume that a polytropic relation with index $\Gamma=4/3$ holds. 
For a radiation dominated flow, $P=a_R T^4/3=K \rho^{4/3}$ ($K$ polytropic constant) and hence 
\begin{equation}
\label{eq:rho}
\rho(r(\mu),t=0) = \rho_0 \times \left[ \frac{sin(\pi x)}{\pi x}\right]^{\frac{3}{4}}\,\mbox{,}
\end{equation}
with $\rho_0=(a_R/3K)^{3/4} T_0^3$. 

Considering that the envelope is expanding homologously, the initial velocity profile is given by:
\begin{equation}
\label{eq:homol_exp}
v(r(\mu),t=0) = V_0 \times \frac{r(\mu)}{R_0}\,\mbox{,}
\end{equation}
where $V_0$ is the initial velocity at the outer radial boundary $R_0$. 

As for the radiation field, in the aforementioned assumptions the diffusion approximation holds. Therefore, $w_0=a_R T^4$ 
and $w_1 \approx w_0/(k_{R}\rho r)$, where $k_{R}$ is approximated with the electron scattering opacity $k_{es}$ 
\citep*[see][]{nobili91}.

For a fixed composition and radial distribution of \chem{56}{Ni}, four input parameters determine uniquely 
the dynamical and thermal properties of the expanding CC-SN envelope (including its initial kinetic and 
thermal energy): the total mass $M_{env}$, the initial radius $R_0$, the initial sound speed at the inner 
boundary $c_{s_0}$, and the ratio $\tilde{k}$ of the initial accretion timescale $t_{acc,0}$ to the initial 
expansion time $t_{exp,0}$, defined as follows \citep*[][]{colpi96,zampieri98}:
\begin{equation}
\label{eq:kappa}
\tilde{k}= \frac{t_{acc,0}}{t_{exp,0}}= \frac{GM_c}{c_{s_0}^3} \times \frac{V_0}{R_0}\,\mbox{,}
\end{equation}
where $M_c$ is the mass of the compact remnant, taken to be equal to $3$\Msun\, at the onset of the 
evolution\footnote{This value of $M_c$ is larger than that of a typical compact remnant ensuing from CC-SN 
events ($\sim1.3$-$2.6$\Msun; see e.g.~\citealt*{limongi00}), but we chose it because it helps the numerical 
stability of the innermost part of the flow. While the dynamical evolution of the expanding ejecta is essentially 
unaffected by $M_c$, the amount of material that falls back in the compact remnant during the evolution is slightly 
overestimated by a factor $\lesssim 3$-$8$\%. Clearly, detailed calculations of fallback require choosing the value 
of $M_c$ more accurately.}. In particular, the polytropic constant $K$ and the initial central temperature $T_0$ 
depend on the three input parameters $c_{s_0}$, $R_0$ and $M_{env}$, and can be written as $K=3c_{s_0}^2/4 \rho_0^{1/3}$ 
and $T_0=(3K/a_R)^{1/4} \rho_0^{1/3}$, where $\rho_0$ is a function of $R_0$ and $M_{env}$. $V_0$ can be expressed 
in terms of $c_{s_0}$, $R_0$ and $\tilde{k}$ through Eq.~[\ref{eq:kappa}]. The dynamical interplay of the inner 
accreting envelope and the outer expanding ejecta can be characterized, in absence of \chem{56}{Ni} energy input, in 
terms of the sole parameter $\tilde{k}$ \citep[see][for details]{colpi96,zampieri98}, while the initial kinetic and 
thermal energy of the envelope are obtained by integration.

Concerning the boundary conditions and their numerical implementation, we refer to \citet{zampieri98}. However we recall 
that at the inner boundary we assume negligible pressure and radiative forces on the gas, and adopt an outgoing wave 
(floating) boundary for $w_1$. At every time step we also set the mass of the compact remnant equal to its initial mass 
plus the actual mass that has fallen back up to that time. At the outer boundary, we adopt free expansion for 
the ejecta, a non-illuminated atmosphere in radiative equilibrium for the radiation field, and synchronization of the 
coordinate time with the proper time of an observer comoving with the ejecta.

\subsection{Input physics}
\label{sec:code:physics}

While a complete description of the input physics adopted in the paper is reported in 
\citet{balberg00}, here we summarize the most relevant aspects concerning the treatment 
of the Rosseland and Planck opacities, the ionization fractions, the gas equations of state, 
and the radioactive heating.

As for the opacity and ionization fractions, the code uses the TOPS opacities available at 
the LANL server \citep[version preceding April 2000; see][]{magee95}, extended in the low-temperature 
regime (T$<$ 5.8$\times$10$^3$ K) using the tables of \citet[][]{AF94}. They are calculated with linear 
interpolation in temperature, density and chemical composition, considering oxygen as representative 
of the entire metal component in the ejecta. A more refined approach using the most up-to-date 
opacities and realistic post-explosive chemical composition (computed following the progenitor 
evolution from the main-sequence up to the explosion) will be presented in a forthcoming paper. 
It should be noted that these opacities are computed assuming LTE between matter and radiation, 
while in the SN ejecta these conditions are not strictly met near to the photosphere. Furthermore, the 
effect of the non-thermal excitation and ionization from gamma rays is not included in the calculation of the opacities.
In order to account for it, a so-called ``opacity floor'' (i.e. a minimum value equal to 0.1 cm$^2$ g$^{-1}$) is used
in all the simulations reported in this paper, following a common practice in the hydrodynamical 
modelling of CC-SNe (e.g. \citealt{herzig90}; \citealt*{swartz91}; \citealt{Y04}; \citealt{dessart10}).

Concerning the gas equation of state, it is approximated as an ideal gas of ions and electrons.

\begin{table}[ht]
  \begin{center}
  \caption{Properties of the relevant radioactive isotopes\label{tab:features}}
  \begin{tabular}{c|ccc}
  Isotope       & $\varepsilon_{\gamma}$  & $\varepsilon_{e^{+}}$   & $\tau$ \\
                & [erg g$^{-1}$ s$^{-1}$] & [erg  g$^{-1}$s$^{-1}$] & [days] \\
  \tableline
  \chem{56}{Ni} &  3.90$\times$10$^{10}$  & 0                       & 8.8    \\
  \chem{56}{Co} &  6.40$\times$10$^{9}$   & 2.24$\times$10$^{8}$    & 111.3  \\
  \chem{57}{Co} &  6.81$\times$10$^{8}$   & 0                       & 391.0  \\
  \chem{44}{Ti} &  2.06$\times$10$^{8}$   & 6.54$\times$10$^{7}$    & 3.28$\times$10$^{4}$\\
  \end{tabular}
  \tablecomments{Data taken from \citet*[][]{woosley89} and \citet*[][]{shigenomoto90}.}
\end{center}
\end{table}

As far as radioactive heating is concerned, we assume that the energy released by
the decays of all the radioactive isotopes present in a given mass shell is absorbed
locally or escapes from the envelope. Non-local effects induced by gamma-ray heating
are not considered. This is implemented adding to the gas energy equation (Eq.~[\ref{eq:energy}]) 
a local energy source term $Q$, which represents the radioactive heating rate of all the
relevant isotopes synthesized in the CC-SN explosion and is given by:
\begin{equation}
\label{eq:dacay}
Q=\sum_i X_i \times (\varepsilon_{i,\gamma} f_i + \varepsilon_{i,e^{+}}) \times e^{-\frac{t}{\tau_i}}\, ,
\end{equation}
where $X_i$, $\tau_i$, $\varepsilon_{i,\gamma}$ and $\varepsilon_{i,e^{+}}$ 
represent the mass fraction, lifetime, energy per unit mass and unit time released by the decay of 
the $i$-th isotope in the form of $\gamma$-rays and positrons, respectively, and the factor $f_i$ 
accounts for the fact that the $\gamma$-rays are not totally trapped in the envelope
\citep[for details, see][and references therein]{balberg00}. The relevant radioactive
nuclei considered in the present investigation and the values of their characteristic parameters are 
reported in Table \ref{tab:features}. The \chem{56}{Co} abundance is evaluated considering that 
the isotope \chem{56}{Co} is involved into the nuclear decay chain \chem{56}{Ni} 
$\rightarrow$ \chem{56}{Co} $\rightarrow$ \chem{56}{Fe}, so its mass as a function of the 
time $t$ is determined from:
\begin{eqnarray}
 \label{eq:56co}
 M_{\mathrm{^{56}{Co}}}(t) = M_{\mathrm{^{56}{Co}},0}\times e^{-\frac{t}{\tau_{\mathrm{^{56}{Co}}}}} + \nonumber \\
 M_{\mathrm{^{56}{Ni}},0} \times \frac{1}{1-\frac{\tau_{\mathrm{^{56}{Ni}}}}{\tau_{\mathrm{^{56}{Co}}}}} \left[e^{-\frac{t}{\tau_{\mathrm{^{56}{Co}}}}}- e^{-\frac{t}{\tau_{\mathrm{^{56}{Ni}}}}}\right] \nonumber \\
 \simeq M_{\mathrm{^{56}{Ni}},0}\times \left[e^{-\frac{t}{\tau_{\mathrm{^{56}{Co}}}}}- e^{-\frac{t}{\tau_{\mathrm{^{56}{Ni}}}}}\right]\, ,
\end{eqnarray}
where we assume that the initial abundance of \chem{56}{Co} is equal to zero and that
$1/(1-\tau_{\mathrm{^{56}{Ni}}}/\tau_{\mathrm{^{56}{Co}}}) \simeq 1$.

\section{Numerical computations}
\label{sec:simul}

\subsection{Preliminary tests}
\label{sec:simul:test}

In order to test the dependability of the new version of the code, a few simulations have been 
performed with both the new and old version. The evolution starts from the same set of initial 
conditions reported in sect.~\ref{sec:code:IniCond}, except for the radial distribution of 
\chem{56}{Ni} that is assumed to be uniform. We varied several input parameters (number of points, 
location of the inner boundary, mass of the ejecta, initial radius and energy) to check the 
stability and accuracy of the code. In the outer expanding part of the flow (which comprises most 
of the mass, in practice all the envelope mass apart from the innermost $\sim 10^{-3}$\msun) the fractional 
difference between the values of the variables computed with the new and old version of the code is 
$\lesssim$10-20\% (e.g., Figures \ref{fig:tvrho} and \ref{fig:w01}; see also \citealt*{pumo09a}).
The only exception is the profile of the radiative flux that differs by $\lesssim 50$\%. Indeed, the radiative 
flux depends sensitively on the numerical treatment. The improved stability of the new version makes 
the profile computed by the new code more reliable. A significant difference ($\lesssim 40$\%) is present 
also in the innermost $\sim 10^{-3}$\msun\, of the envelope (below $\log R \sim 12.7$), where the flow starts 
to fall back onto the remnant. This is most likely linked to the delicate balance between the pressure 
gradients and the gravitational force, that determines the location of the accretion radius $r_a$ 
\citep[see][for details]{balberg00}. Even a slight difference in the calculation of the gas and radiation 
pressures may cause a sign reversal in the velocity of the marginally bound gas shells (containing 
$\lesssim 10^{-4}$\msun\,) located near to $r_a$, that start then to fall back onto the remnant instead of 
going outwards.

\placefigure{fig:tvrho}
\begin{figure}[h]
\epsscale{0.82}
\begin{turn}{-90}
\plotone{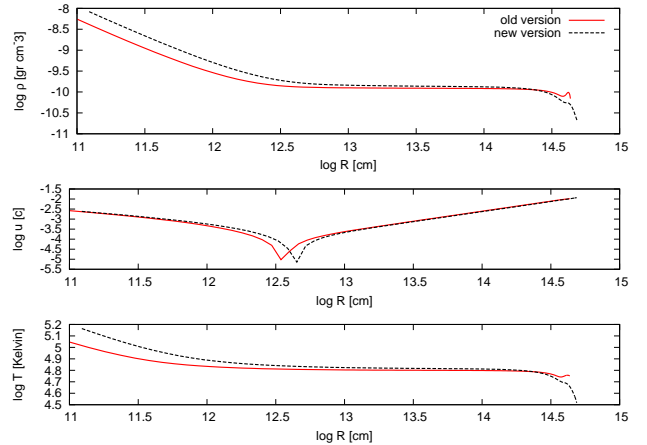}
\end{turn}
\caption{Comparison between results obtained from the old (dotted line) and new (dashed line) 
versions of the code. The comparison refers to a model evolved for 15 days from the breakout 
of the shock wave at the stellar surface, and having initial radius of $3\times10^{13}$ cm, 
total energy of 1 foe, amount of \chem{56}{Ni} equal to 0.07 \Msun, and the envelope mass of 
16 \Msun. Top, middle, and bottom panels show the radial profiles of the gas matter density 
$\rho$ (in units of g cm$^{-3}$), absolute value of the fluid radial-velocity (in units of 
the velocity of light c) and the gas temperature (in units of Kelvin), respectively. The 
innermost part of the envelope, below $\log R \sim 12.5$-$12.7$, is accreting onto the central
remnant and has negative velocity. \label{fig:tvrho}}
\end{figure}
\placefigure{fig:w01}
\begin{figure}[h]
\epsscale{0.82}
\begin{turn}{-90}
\plotone{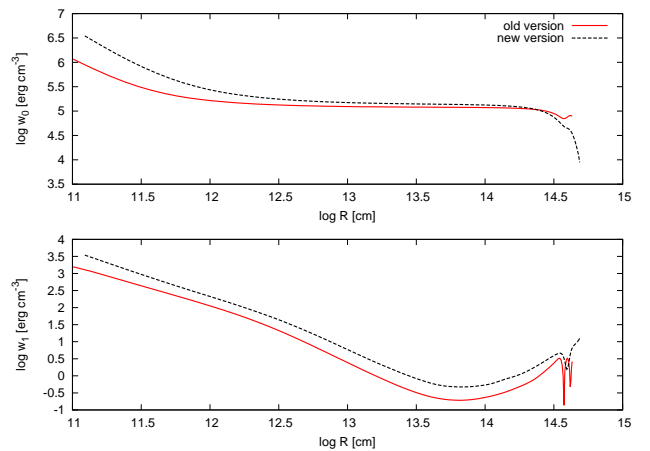}
\end{turn}
\caption{Same as Fig.~\ref{fig:tvrho}, but for the radial profiles of the radiation energy 
density $w_0$ (top panel) and the radiative flux $w_1$ (bottom panel). Both quantities are 
in units of erg cm$^{-3}$. \label{fig:w01}}
\end{figure}
\placefigure{fig:RecombFront}
\begin{figure}[h]
\epsscale{0.82}
\begin{turn}{-90}
\plotone{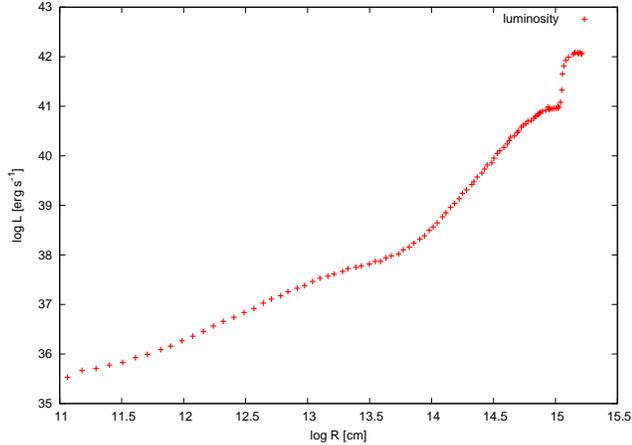}
\end{turn}
\caption{Radial profiles of the radiative luminosity at the beginning of the recombination 
phase for model (1) in Table \ref{tab_models}. Less than 5\% in mass of the envelope has 
recombined. The sharp front at $\log R \sim 15.1$ marks the position of the recombination 
wave, where also the photosphere is located.\label{fig:RecombFront}}
\end{figure}

While the simulations with the old version often develop numerical instabilities, those evolved 
with the new version show to be stable. Several models were evolved for long times checking the 
long-term numerical stability of the new code during some key evolutionary stages like the 
radiative recombination phase. The position of the recombination front is very well traced (see 
Figure~\ref{fig:RecombFront}) and, consequently, it is possible to follow the post-explosion 
evolution of a CC-SN event in its ``entirety'', from the breakout of the shock wave at the stellar 
surface up to the nebular stage.

\subsection{Grid of models}
\label{sec:simul:models}

Different simulations were performed with the new version of the code exploring a relatively large 
fraction of the parameter space related to the post-explosion evolution of a ``typical'' CC-SN.

In particular a total of 22 models were calculated changing the initial radius from $3\times10^{12}$ 
to $10^{14}$ cm, the total energy from 0.5 to 3 foe, the amount of \chem{56}{Ni} from 0.005 to 0.070 
\Msun, and the envelope mass from 8 to 18 \Msun (see Table \ref{tab_models} for details). 
In the present sample, we assume that 
the initial thermal and kinetic energy are equal and hence three out of the four input parameters 
of the model ($M_{env}$, $c_{s_0}$ and $\tilde{k}$) are related.
All the simulations were performed fixing the parameter controlling the \chem{56}{Ni} distribution at the 
value $K_{mix}=5$. This corresponds to having $\sim$ 90\% of the \chem{56}{Ni} in the envelope 
concentrated in the innermost $\sim$ 3, 5, 6.5 and 7.5\Msun\, for the models with an envelope mass 
equal to 8, 12, 16 and 18\Msun, respectively (see Eq.~[\ref{eq:Ni56}]).

\begin{table}[ht]
  \begin{center}
  \caption{Models Parameters\label{tab_models}}
  \begin{tabular}{c|cccc}
  \tableline
  \tableline
  Model & Radius        & Envelope mass & Energy                          & \chem{56}{Ni} mass\\
        & 10$^{12}[cm]$ & $[M_\odot]$   & [foe$\equiv$10$^{51}$\,ergs)]   & $[M_\odot]$       \\
  \tableline
  1     & $3$   & 16  & 1   & 0.070  \\ 
  2     & $30$  & 18  & 1   & 0.070  \\ 
  3     & $30$  & 8   & 1   & 0.035  \\ 
  4     & $30$  & 12  & 1   & 0.035  \\ 
  5     & $30$  & 12  & 2   & 0.035  \\ 
  6     & $30$  & 16  & 0.5 & 0.070  \\
  7     & $30$  & 16  & 1   & 0.005  \\ 
  8     & $30$  & 16  & 1   & 0.010  \\ 
  9     & $30$  & 16  & 1   & 0.020  \\ 
  10    & $30$  & 16  & 1   & 0.035  \\ 
  11    & $30$  & 16  & 1   & 0.045  \\ 
  12    & $30$  & 16  & 1   & 0.055  \\ 
  13    & $30$  & 16  & 1   & 0.070  \\ 
  14    & $30$  & 16  & 1.3 & 0.070  \\ 
  15    & $30$  & 16  & 1.6 & 0.070  \\ 
  16    & $30$  & 16  & 1.9 & 0.070  \\ 
  17    & $30$  & 16  & 2   & 0.035  \\ 
  18    & $30$  & 16  & 2   & 0.070  \\ 
  19    & $30$  & 16  & 3   & 0.035  \\ 
  20    & $30$  & 16  & 3   & 0.070  \\ 
  21    & $60$  & 16  & 1   & 0.070  \\ 
  22    & $100$ & 16  & 1   & 0.070  \\ 
  \end{tabular}
\end{center}
\end{table}

\section{Validation of the new code and physical behavior of CC-SN models}
\label{sec:validation}

\subsection{Code validation}

Some of the models reported in Table \ref{tab_models} have been used to further validate the 
code against observations and by comparison with similar models reported in the literature. 
In particular, we have compared:
\begin{itemize}
 \item[-] model (1) with model (A) of \citet{Y04} that has a similar input parameters (i.e. total 
 energy of 1 foe, initial radius of $3\times10^{12}$ cm, \chem{56}{Ni} mass of 0.07\Msun, and 
 \chem{56}{Ni} mixing throughout the innermost 6 \Msun\, region of the envelope);
 \item[-] models (3), (4) and (10) with results of simulations performed with the semi-analytic 
 code of \citet{zampieri03} and \citet{zampieri07};
 \item[-] model (13) with results of simulations performed with the aforementioned semi-analytic 
 code, and with model (B) of \citet{Y04} that has a similar input physics (i.e. total energy of 1 
 foe, initial radius of $3\times10^{13}$ cm, \chem{56}{Ni} mass of 0.07\Msun, and \chem{56}{Ni} 
 mixing throughout the innermost 6 \Msun\, region of the envelope);
 \item[-] models (1) and (2) with observations of SN 1987A.
\end{itemize}

Figures \ref{fig:mod1vsY} through \ref{fig:VT87A} show the results of these comparisons.

\placefigure{fig:mod1vsY}
\begin{figure}[h]
\epsscale{0.82}
\begin{turn}{-90}
\plotone{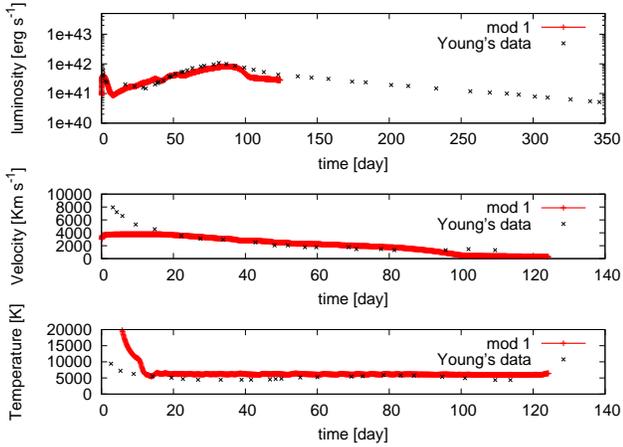}
\end{turn}
\caption{Evolution of the main observables for model (1) in Table \ref{tab_models} and model 
(A) of \citet{Y04} (see text for details). Top, middle, and bottom panels show the bolometric 
light curve, photospheric velocity and temperature as a function of time, respectively.
\label{fig:mod1vsY}}
\end{figure}

\placefigure{fig:mod3vsSA}
\begin{figure}[h]
\epsscale{0.82}
\begin{turn}{-90}
\plotone{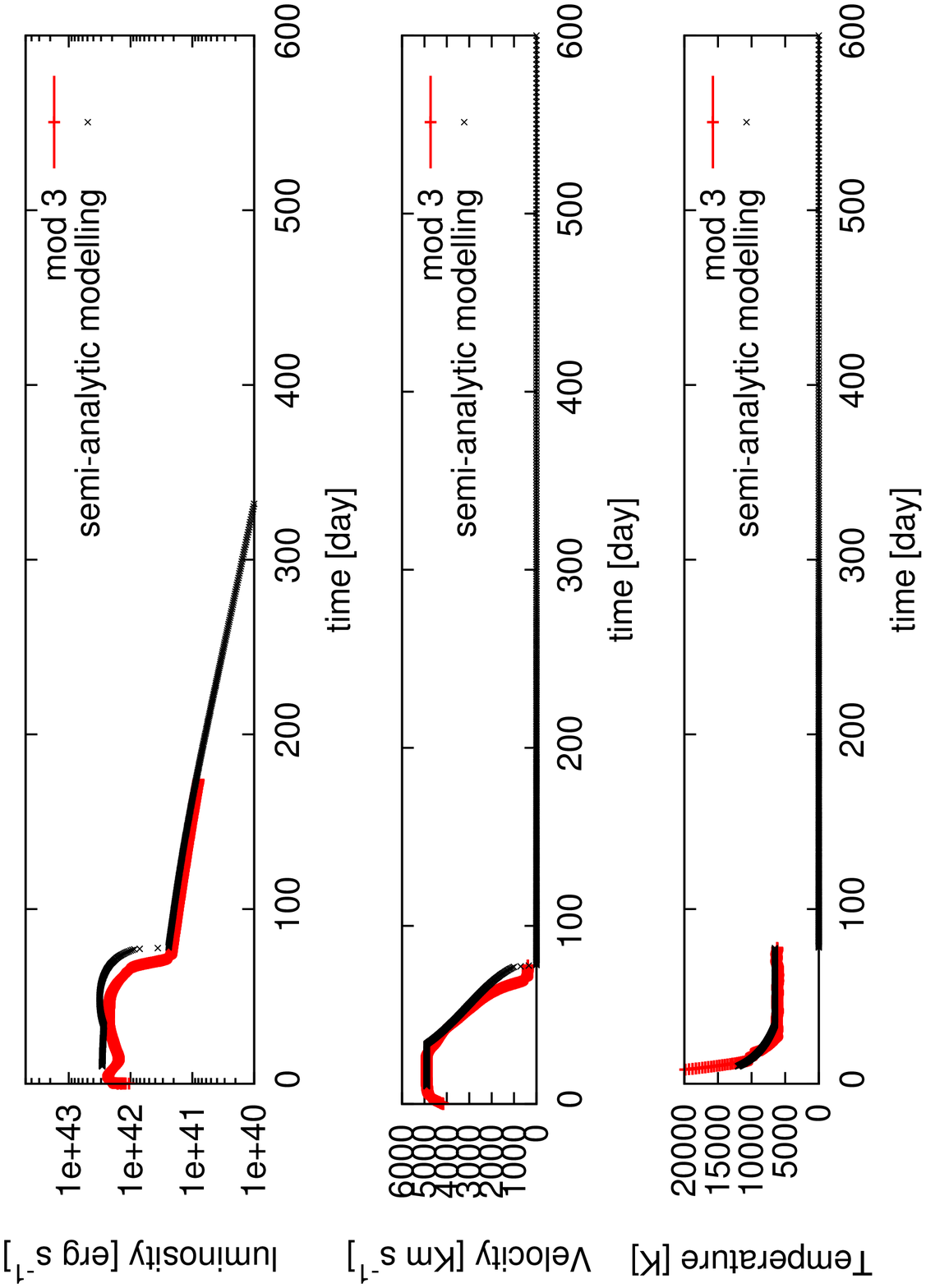}
\end{turn}
\caption{Same as Fig.~\ref{fig:mod1vsY}, but for the comparison between model (3) and 
the corresponding model computed with the semi-analytic code (see text for details).
\label{fig:mod3vsSA}}
\end{figure}

\placefigure{fig:mod4vsSA}
\begin{figure}[h]
\epsscale{0.82}
\begin{turn}{-90}
\plotone{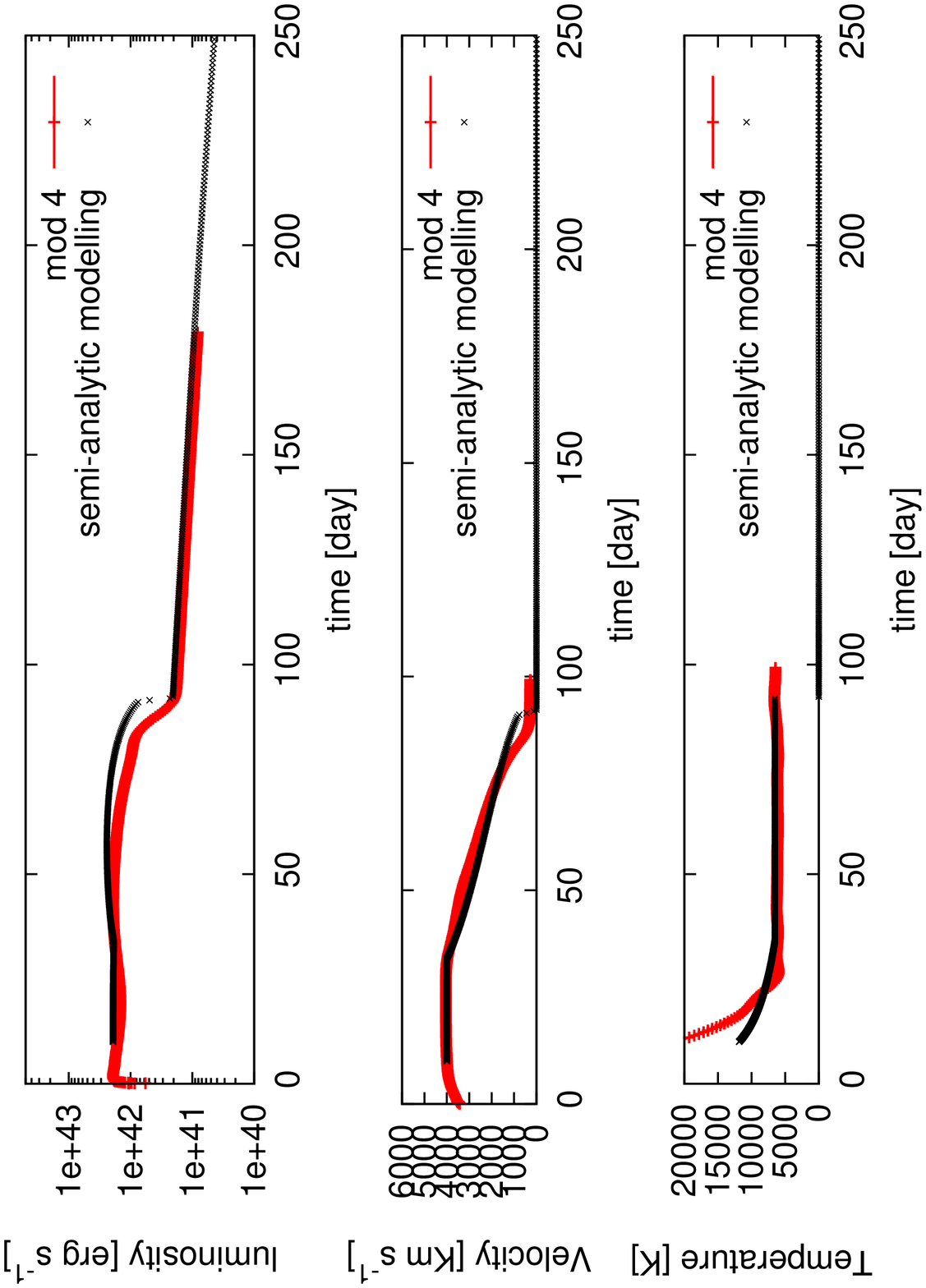}
\end{turn}
\caption{Same as Fig.~\ref{fig:mod1vsY}, but for the comparison between model (4) and 
the corresponding model computed with the semi-analytic code (see text for details).
\label{fig:mod4vsSA}}
\end{figure}

\placefigure{fig:mod10vsSA}
\begin{figure}[ht]
\epsscale{0.82}
\begin{turn}{-90}
\plotone{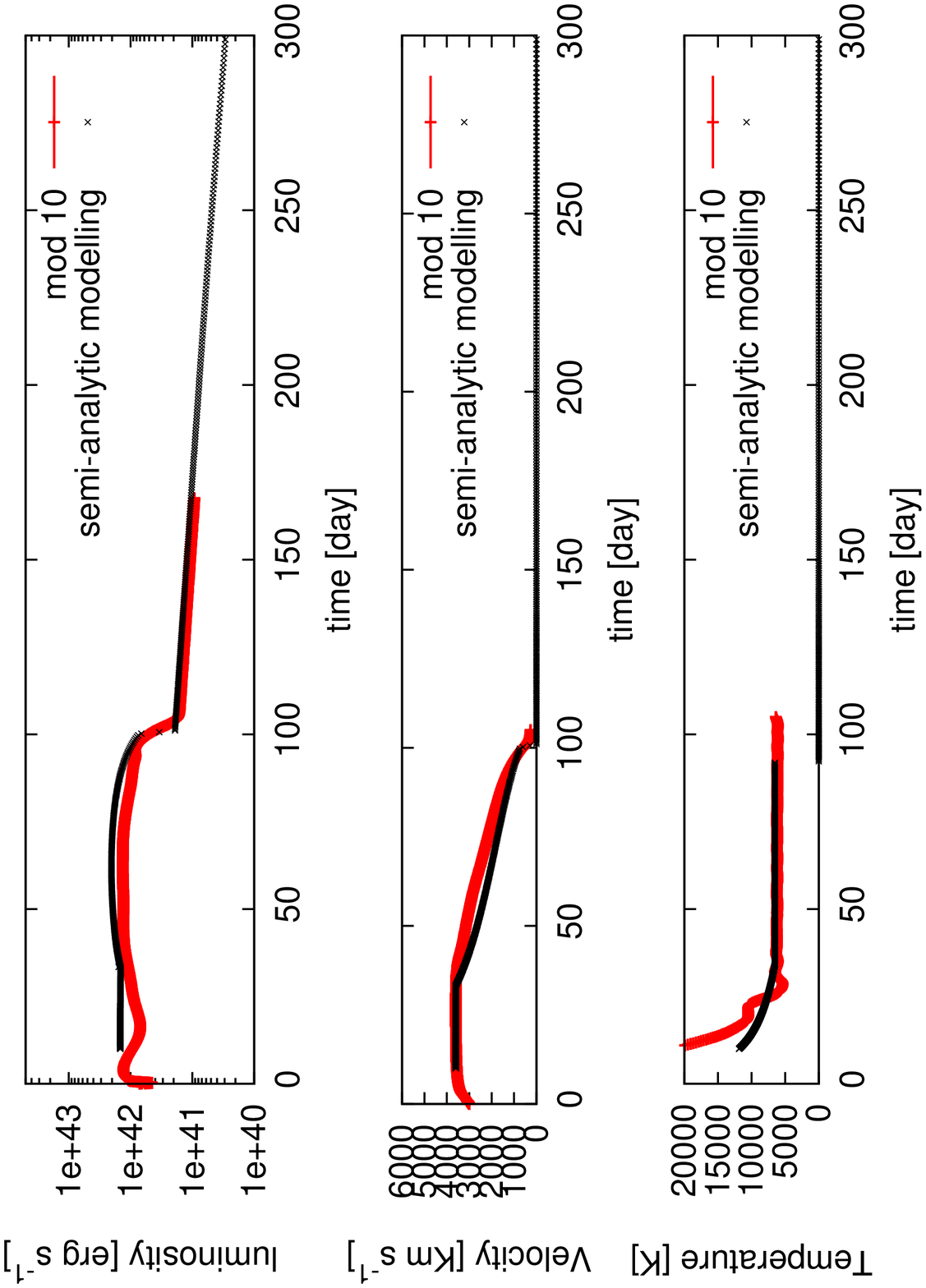}
\end{turn}
\caption{Same as Fig.~\ref{fig:mod1vsY}, but for the comparison between model (10) and 
the corresponding model computed with the semi-analytic code (see text for details).
\label{fig:mod10vsSA}}
\end{figure}

\placefigure{fig:mod13vsSA}
\begin{figure}[h]
\epsscale{0.82}
\begin{turn}{-90}
\plotone{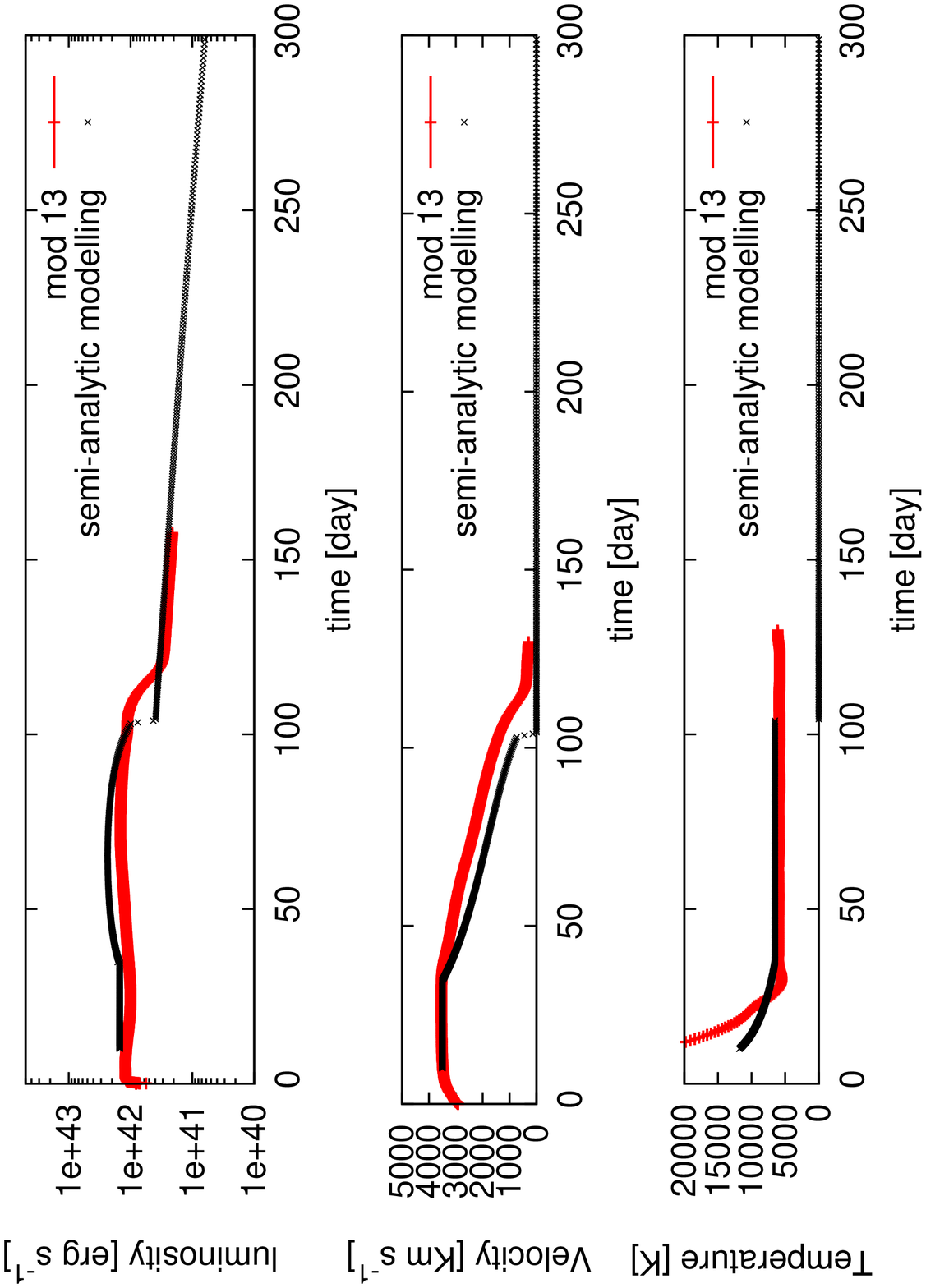}
\end{turn}
\caption{Same as Fig.~\ref{fig:mod1vsY}, but for the comparison between model (13) and 
the corresponding model computed with the semi-analytic code (see text for details).
\label{fig:mod13vsSA}}
\end{figure}

\placefigure{fig:mod13vsY}
\begin{figure}[h]
\epsscale{0.82}
\begin{turn}{-90}
\plotone{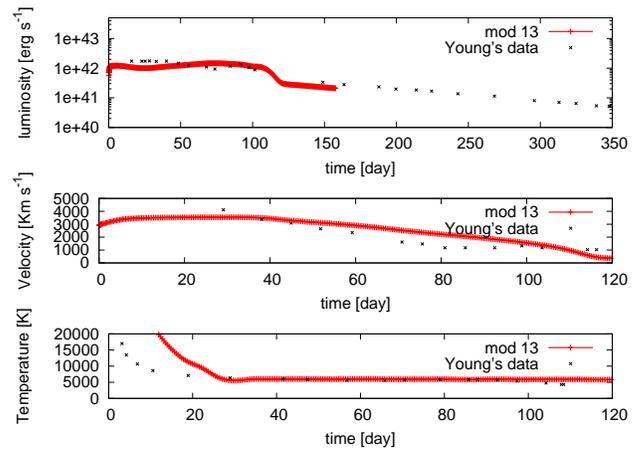}
\end{turn}
\caption{Same as Fig.~\ref{fig:mod1vsY}, but for the comparison between our model (13) 
and model (B) of \citet{Y04} (see text for details).\label{fig:mod13vsY}}
\end{figure}

\placefigure{fig:LC87A}
\begin{figure}[h]
\epsscale{0.82}
\begin{turn}{-90}
\plotone{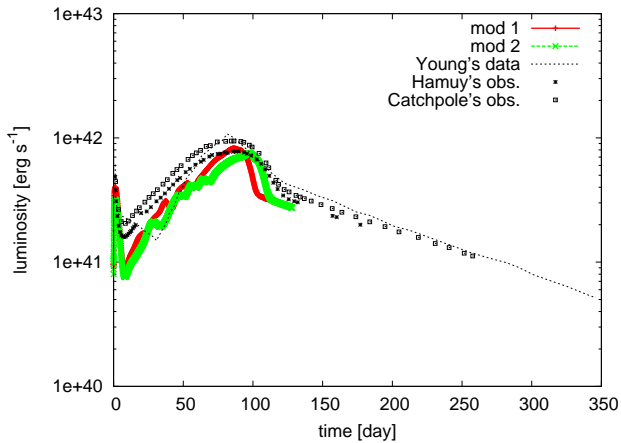}
\end{turn}
\caption{Bolometric light curves of models (1) and (2) compared to the light curve of 
SN 1987A (taken from \citealt{Catchpole87,Catchpole88} and \citealt{hamuy88}). The best 
fitting model of \citet{Y04} is also reported for comparison.\label{fig:LC87A}}
\end{figure}

\placefigure{fig:VT87A}
\begin{figure}[h]
\epsscale{0.82}
\begin{turn}{-90}
\plotone{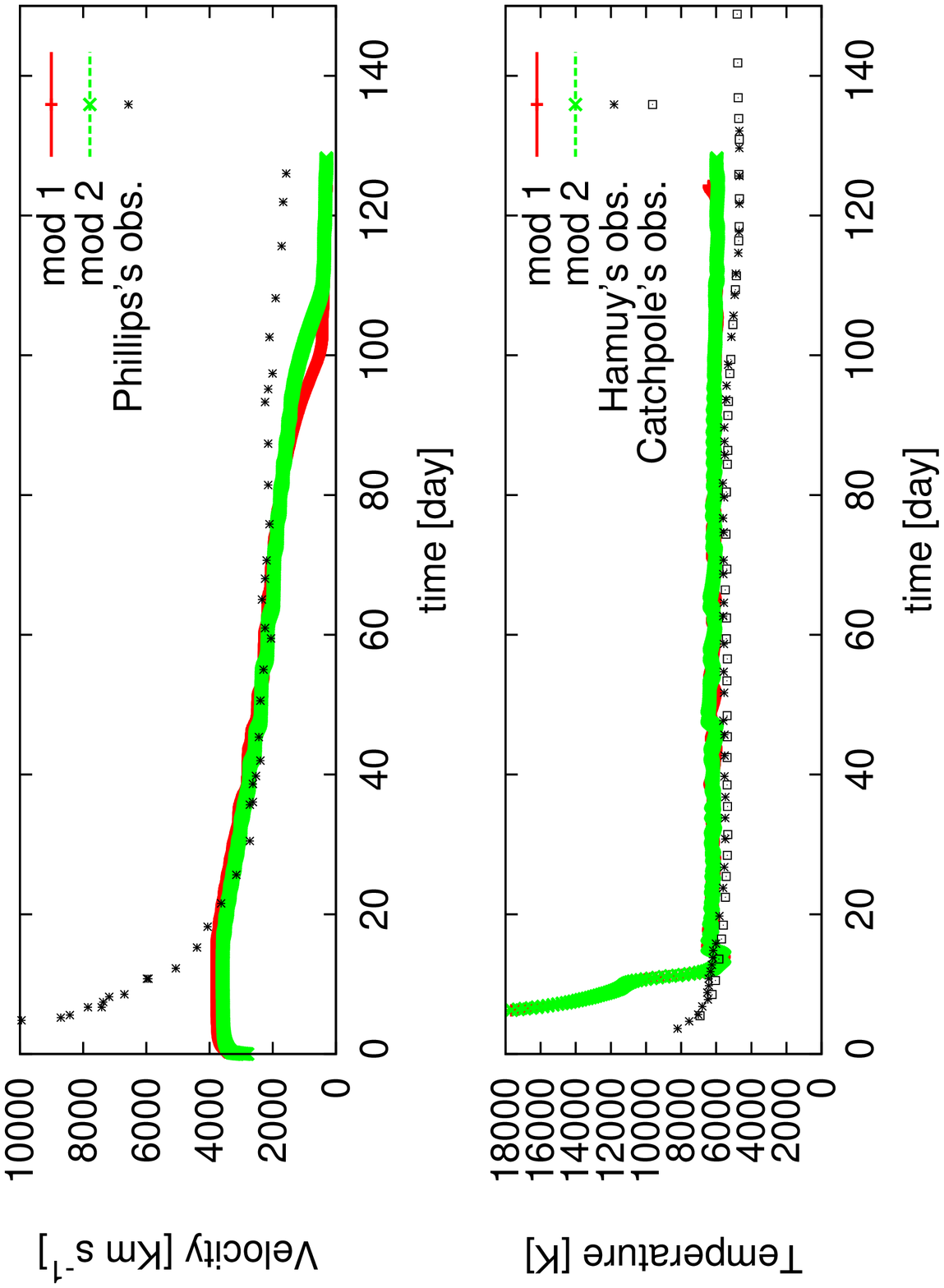}
\end{turn}
\caption{Evolution of the photospheric velocity (top panel) and temperature (bottom panel) 
for models (1) and (2) compared to the corresponding quantities of SN 1987A (taken from 
\citealt{Catchpole87,Catchpole88}, \citealt{hamuy88} and \citealt{phillips88}).\label{fig:VT87A}}
\end{figure}

The agreement between the results obtained with our code and those reported in the literature 
is good, considering the different initial conditions, input physics and numerical methods 
adopted (see Figures \ref{fig:mod1vsY} through \ref{fig:mod13vsY}). 
In particular, we found overall consistency with the results of the semi-analytic code of 
\citet{zampieri03} and \citet{zampieri07}, that serves also as cross validation of the latter.

However, in some simulations there are somewhat significant discrepancies in the photospheric velocity 
and temperature during the early evolution ($\la$ 10-20 days). The fractional difference between the 
values of the variables computed with our new code and those computed with other codes is $\ga$ 15-30\%.
In particular, the disagreement with the models computed by \citet{Y04} is probably due to the initial 
density profile. Indeed, in this investigation we limited our analysis to simplified initial conditions, 
assuming that gas behaves as a polytrope and that $\rho \propto T^{3} \propto [sin(\pi x)/(\pi x)]^{3/4}$ 
(see Eq.~[\ref{eq:rho}]). However, the external layers of a SN typically have a steeper power-law distribution 
caused by the acceleration of the shock wave through the exponentially decaying density profile of the 
progenitor star. These layers (0.01-0.1 \Msun\, in mass) are set into high-velocity and are not homologously 
expanding \citep[see also][]{utrobin04,woosley88}. During the first $\sim$10-20 days after the explosion, the 
photosphere is located in this shell, which is then cooler and has a higher velocity than the bulk of the 
ejecta. This results in a different evolution of the photospheric temperature and velocity. 
A better agreement during this phase is found with the semi-analytic code, that adopts a uniform density 
distribution. Some disagreement is present also at late-times ($\gtrsim$ 80-110 days), that is probably 
related to differences in the treatment of ionization balance and in the adopted opacities (including the 
opacity floor).

Despite the limitations of our simplified initial conditions, also the comparison with the bolometric 
light curve of SN 1987A is satisfactory. We are able to reproduce its main features (peak luminosity 
and phase at maximum) with models having initial radius of $3\times10^{12}$ cm, total initial 
energy of $1$ foe, amount of \chem{56}{Ni} equal to $0.07$\Msun, and envelope mass ranging between 
16 and 18 \Msun\, (models (1) and (2) in Figures \ref{fig:LC87A} and \ref{fig:VT87A}). 
The agreement can be considered satisfactory also because we did not perform any ``fine-tuning'' of 
the initial composition which is typically needed in order to accurately reproduce the observed shape 
(width and rise to peak) of the bolometric light curve of SN 1987A \citep[e.g.][]{woosley88,utrobin04}. 
Further residual differences may also be caused by the absence of non-thermal ionization from gamma 
rays in our models. The time evolution of the photospheric velocity and temperature of SN 1987A is also 
well reproduced by models (1) and (2), apart from the differences in the early and the late-time evolution 
due to the same reasons mentioned above. Moreover, as can be seen from Figure \ref{fig:LC87A}, the luminosity in the 
radioactive tail predicted by the model is lower than the observed one. This is a consequence of fallback 
occurring during the evolution. We found that the innermost $\sim0.01$\Msun\, of the envelope, 
containing $\sim 2.4\times10^{-3}$\Msun\, of \chem{56}{Ni}, have been accreted onto the central remnant.

\subsection{Reference case}

To give a general overview of the post-explosion evolution of a ``typical'' CC-SN, we focus on the properties 
of our model (13) that has rather common initial parameters. The evolution is determined by the thermodynamics
of the expanding ejecta. The internal energy deposited by the shock wave and that released by gamma-ray 
radioactive decays are used to expand the ejecta and power light curve. The evolution is characterized by three 
phases in which different heating and emission mechanisms dominate. During the first phase (diffusive phase), 
the envelope is completely ionized and optically thick, and the emission is due to the release of internal energy 
on a diffusion timescale. In the second phase (recombination phase), the ejecta are recombining and the emission 
is dominated by the sudden release of energy caused by the receding motion of the wavefront through the envelope. 
During the last phase (radioactive-decay phase or radioactive tail), the envelope is recombined and optically thin 
to optical photons, and the emission comes from the thermalization of the energy deposited by gamma-ray photons. 
Observationally, the first two phases coincide with what is usually defined the plateau or photopheric phase, while 
the last phase is referred to as nebular phase.

Figures \ref{fig:modB_18d} through \ref{fig:modB_130d} show the physical properties of model (13) at three different 
times, taken to be representative of the aforementioned three phases. Moreover Figures \ref{fig:modB_TvsR} through 
\ref{fig:modB_radius} show the evolution of the photopheric radius, and the radial profiles of the photospheric 
velocity and temperature in more detail.

During the first phase ($\la 35$ days), the radiation diffusion time-scale is much longer than the expansion time-scale, 
and the cooling induced by photon diffusion is negligible. Although the internal energy decreases because of expansion,
the temperature and the density are sufficiently high that the envelope remains completely ionized and optically thick 
(Figures \ref{fig:modB_18d}, \ref{fig:modB_TvsR} and \ref{fig:modB_TvsM}). The photopheric radius, which is located in the 
outermost shell, increases by a factor $\sim$ 40 in $\sim$ 35 days from the breakout of the shock wave at the stellar 
surface (see Figure \ref{fig:modB_radius}). The expansion is nearly homologous, as witnessed by the nearly constant value 
of the photospheric velocity.

\placefigure{fig:modB_18d}
\begin{figure}[h]
\epsscale{0.82}
\begin{turn}{-90}
\plotone{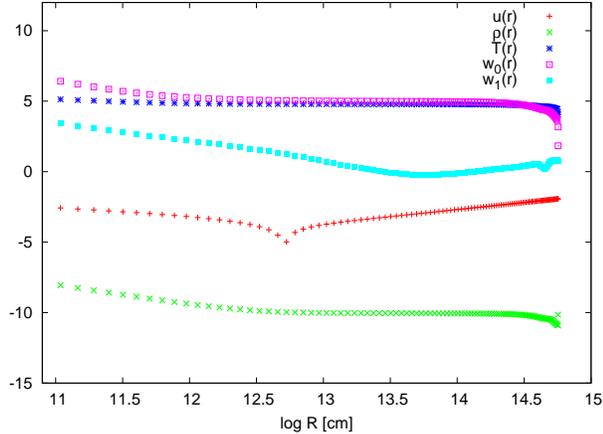}
\end{turn}
\caption{From bottom to top, radial profiles of the gas matter density (in units of g cm$^{-3}$), 
absolute value of the fluid radial-velocity (in units of c), radiative flux (in units of erg cm$^{-3}$), 
gas temperature (in units of Kelvin), and radiation energy density (in units of erg cm$^{-3}$) as 
a function of radius (in units of cm). All scales on the $y$ axis are logarithmic. The radial profiles 
refer to model (13) at 18 days from the breakout of the shock wave at the stellar surface. The 
innermost part of the envelope, below $\log R \sim 12.7$, is accreting onto the central remnant 
and the radial-velocity there is negative. \label{fig:modB_18d}}
\end{figure}

\placefigure{fig:modB_75d}
\begin{figure}[h]
\epsscale{0.82}
\begin{turn}{-90}
\plotone{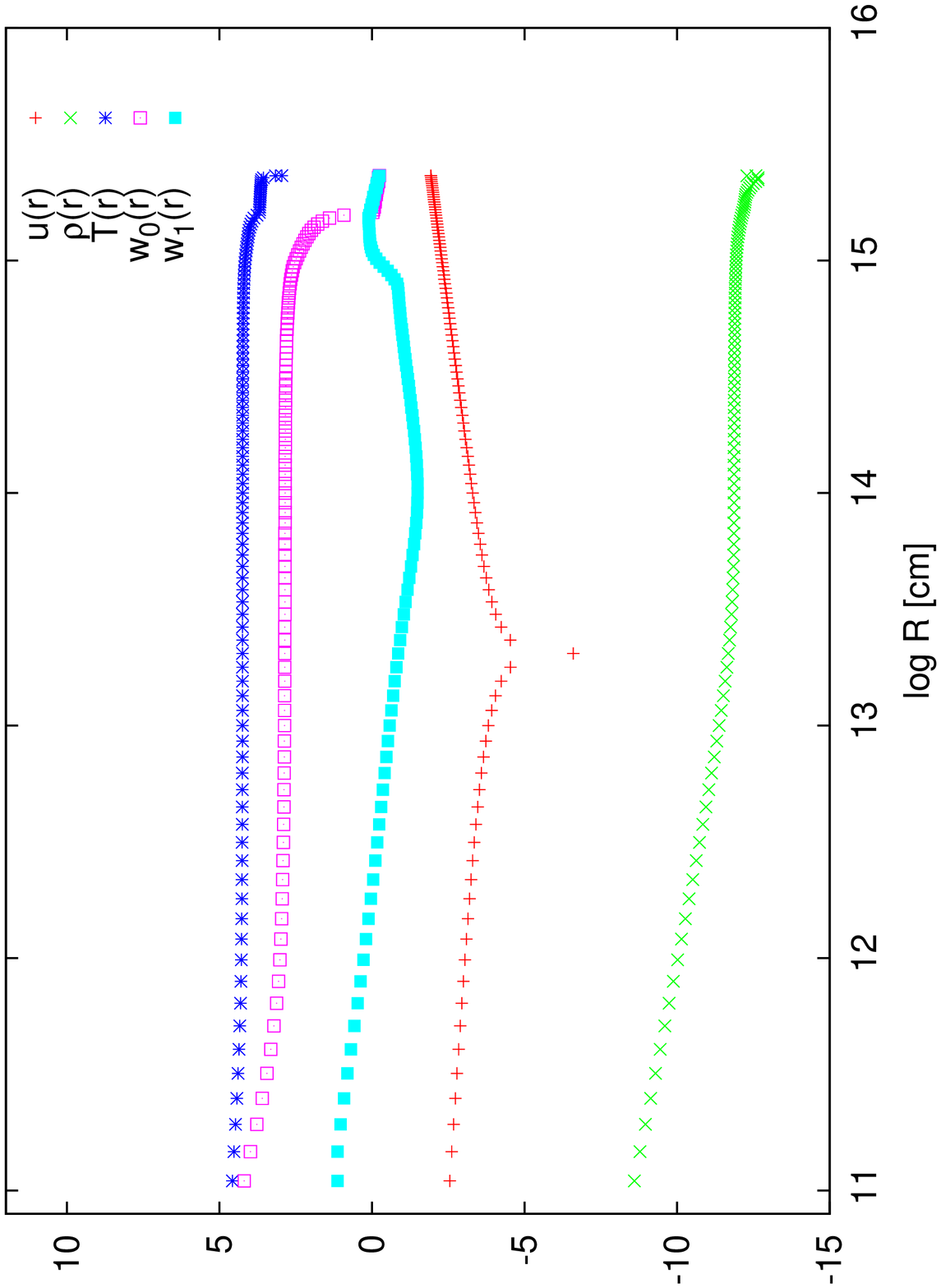}
\end{turn}
\caption{Same as Fig.~\ref{fig:modB_18d}, but for 75 days from the breakout of the shock 
wave at the stellar surface. The boundary of the innermost accreting part of the envelope 
is at $\log R \sim 13.3$. \label{fig:modB_75d}}
\end{figure}

\placefigure{fig:modB_130d}
\begin{figure}[h]
\epsscale{0.82}
\begin{turn}{-90}
\plotone{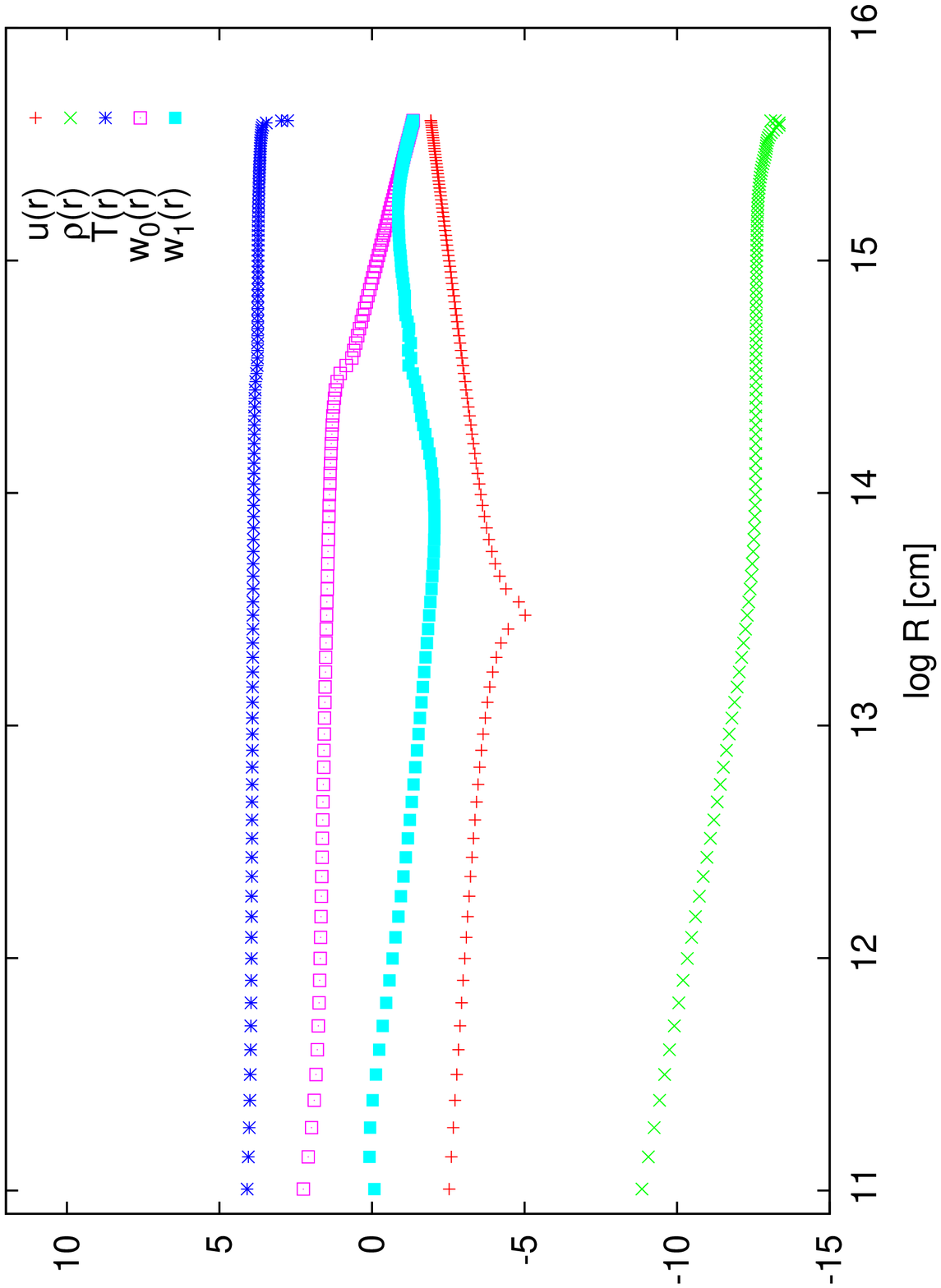}
\end{turn}
\caption{Same as Fig.~\ref{fig:modB_18d}, but for 130 days from the breakout of the shock 
wave at the stellar surface. The boundary of the innermost accreting part of the envelope 
is at $\log R \sim 13.4$. \label{fig:modB_130d}}
\end{figure}

\placefigure{fig:modB_TvsR}
\begin{figure}[h]
\epsscale{0.82}
\begin{turn}{-90}
\plotone{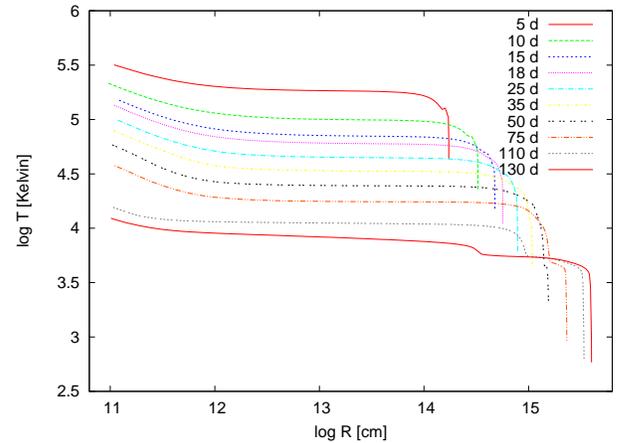}
\end{turn}
\caption{Gas temperature as a function of radius for model (13) at different times. Labels 
indicate the time (in units of days) from the breakout of the shock wave at the stellar 
surface. \label{fig:modB_TvsR}}
\end{figure}

\placefigure{fig:modB_TvsM}
\begin{figure}[h]
\epsscale{0.82}
\begin{turn}{-90}
\plotone{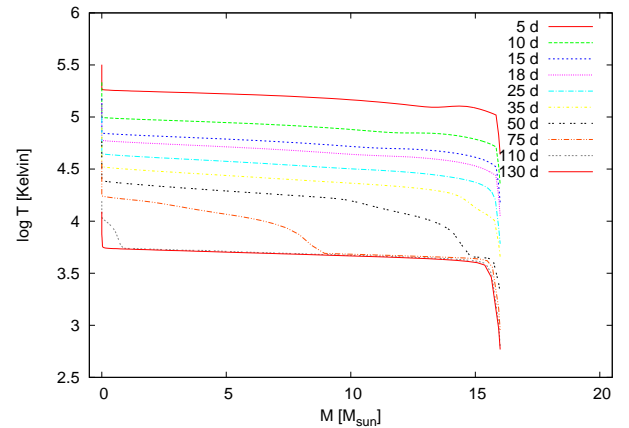}
\end{turn}
\caption{Same as Fig.~\ref{fig:modB_TvsR}, but for the gas temperature as a function of mass.
\label{fig:modB_TvsM}}
\end{figure}

\placefigure{fig:modB_VvsR}
\begin{figure}[h]
\epsscale{0.82}
\begin{turn}{-90}
\plotone{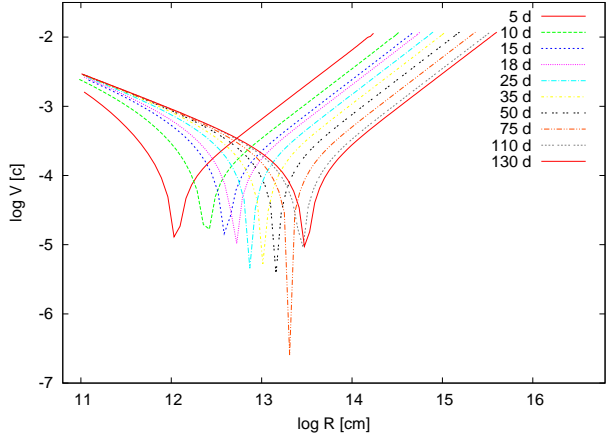}
\end{turn}
\caption{Absolute value of the fluid radial-velocity as a function of radius for model (13) at 
different times. Labels indicate the time (in unit of days) from shock breakout. The minimum of 
each curve marks the boundary of the inner accreting zone (where the radial-velocity is negative).
\label{fig:modB_VvsR}}
\end{figure}

\placefigure{fig:modB_radius}
\begin{figure}[h]
\epsscale{0.82}
\begin{turn}{-90}
\plotone{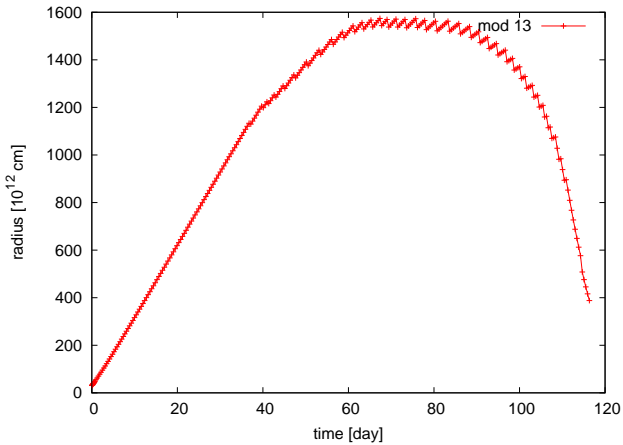}
\end{turn}
\caption{Evolution of the photospheric radius for model (13).\label{fig:modB_radius}}
\end{figure}

At $\sim$ 35 days, in the outermost layers the temperature reaches $\sim$ 6000 K and hydrogen starts 
to recombine. This marks the beginning of the recombination phase. During this phase, the evolution 
is characterized by the fast recombination of the outer zone, leading to the formation of a recombination 
wave (RW) that moves inward (in mass) like a flame. In the external layers, 
the temperature drops below $\sim$ 6000 K and, consequently, the internal energy is efficiently radiated 
away because of the sudden decrease of the opacity (see Figures \ref{fig:modB_75d}, \ref{fig:modB_TvsR} 
and \ref{fig:modB_TvsM}). The RW marks the boundary between the inner envelope that is optically thick, 
ionized and hot ($>$ 6000 K), and the outer layers that are optically thin, recombined and cooler 
($\lesssim$ 6000 K). The photosphere moves inward (in mass) following the motion of the RW and allows photons 
to escape sooner than they would if the photosphere were fixed in the outmost layer of the envelope. As 
a consequence, advective cooling induced by the WR motion becomes dominant. The energy radiated away at this 
stage is the sum of the residual internal energy left over after expansion and that liberated during recombination.
The energy deposited by gamma-rays becomes significant only when the RW reaches the innermost zones, full of 
\chem{56}{Ni} \citep[see also][]{Y04,bersten11}. The bolometric luminosity, initially almost constant, plummets 
when the RW reaches the center (see the top panel in Figure \ref{fig:mod13vsY}). The bulk of the envelope ($\sim$ 
95\% in mass) is recombined after $\sim$ 110 days from the breakout of the shock wave at the stellar surface (see 
Figure \ref{fig:modB_TvsM}). The photopheric velocity decreases because the photosphere moves inward with the RW, 
reaching inner and slower layers as time elapses (see the middle panel in Figure \ref{fig:mod13vsY}). 

The velocity of propagation of the RW is determined by the physical conditions in the envelope, in particular the 
temperature, density, amount of \chem{56}{Ni} and their distributions. The motion of the photospheric radius for a 
distant observer is the result of the interplay between the inward motion of the RW in mass and the expansion of the 
ejecta. Consequently, as shown in Figure \ref{fig:modB_radius}, from $\sim$ 35 up to $\sim$ 65 days, the photospheric 
radius increases, while later on it stays at an almost constant (eulerian) radial coordinate for about $\sim$ 20 days. 
Then, the photosphere starts to recede in radius and the luminosity plummets.

The duration and steepness of the transition from the plateau to the radioactive tail depend on the \chem{56}{Ni} mass 
(see sect. \ref{subsec:survey}) and its distribution in the ejecta. During this last phase, 
the evolution is completely governed by the deposition and reprocessing of the energy released from the radioactive decay 
of \chem{56}{Co} into \chem{56}{Fe}. The envelope is sufficiently opaque to the gamma-rays emitted in the radioactive 
decays that a large fraction of them are effectively absorbed and their energy deposited in the ejecta. This energy is 
easily radiated away by lower energy photons on a timescale much shorter than the expansion timescale. Also the heating 
time, which is essentially the \chem{56}{Co} decay time, is smaller than the expansion time and, hence, the loss of internal 
energy due to expansion is negligible. Therefore, the bolometric luminosity is the total heating rate from the radioactive
decays, proportional to the mass of \chem{56}{Ni} synthesized in the explosion (see also sect.~\ref{subsec:survey}).

\subsection{Effects of varying the initial parameters on the light curve, photospheric velocity and temperature}
\label{subsec:survey}

The sample of models (reported in Table \ref{tab_models})
enabled us to study the effects of the variation of each parameter (namely, envelope mass $M_{env}$, 
initial outer radius $R_0$, total ejecta energy $E$, and amount of \chem{56}{Ni} $M_{Ni}$) on the main 
observables, while holding the others fixed. Figures \ref{fig:mass_env} through \ref{fig:ni} summarize the 
results of this analysis. We note, however, that this investigation is not meant to be a realistic model survey
for a detailed quantitative comparison with observations, because it does not include models computed 
starting from realistic initial conditions. This is postponed to a follow-up investigation.

\placefigure{fig:mass_env}
\begin{figure}[h]
\epsscale{0.82}
\begin{turn}{-90}
\plotone{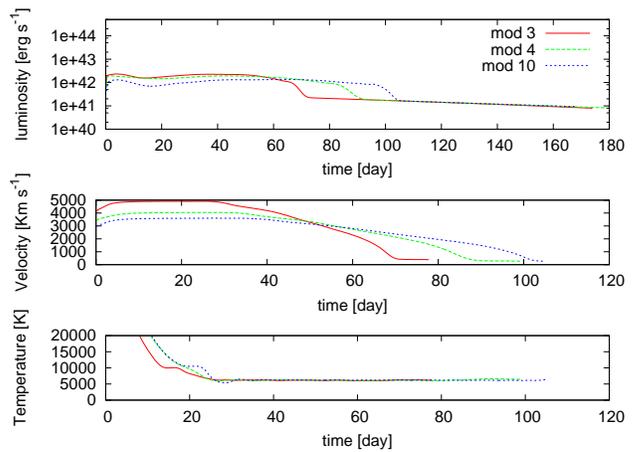}
\end{turn}
\caption{Effects of varying the envelope mass on the main observables for models (3), (4), and
(10) with $M_{env}=8$, $12$, and $16$\Msun, respectively. The models (see Table \ref{tab_models}) 
have the same initial radius ($R_0=3\times 10^{13}$ cm), total energy ($E=1$ foe), and amount of 
\chem{56}{Ni} ($M_{Ni}=0.035$\Msun). Top, middle, and bottom panels show the bolometric light curve, 
photospheric velocity and temperature as a function of time, respectively.
\label{fig:mass_env}}
\end{figure}

\placefigure{fig:radius}
\begin{figure}[h]
\epsscale{0.82}
\begin{turn}{-90}
\plotone{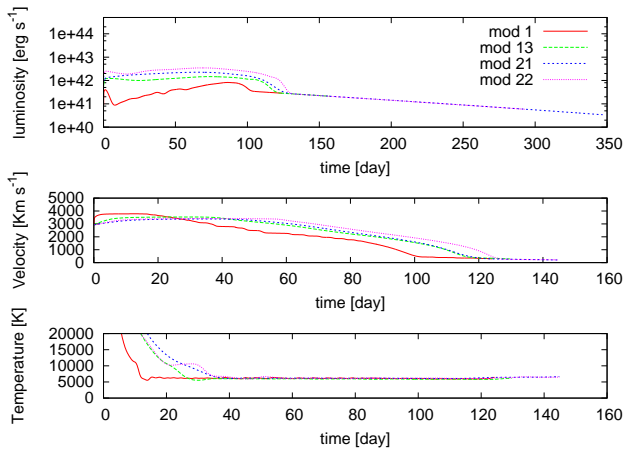}
\end{turn}
\caption{Same as Fig.~\ref{fig:mass_env}, but for the initial radius. The lines refer to models
(1), (13), (21), and (22) with $R_0=3\times10^{12}$, $3\times10^{13}$, $6\times10^{13}$, and $10^{14}$ 
cm, respectively. The models (see Table \ref{tab_models}) have the same envelope mass ($M_{env}=16$\Msun), 
total energy ($E=1$ foe), and amount of \chem{56}{Ni} ($M_{Ni}=0.07$\Msun).
\label{fig:radius}}
\end{figure}

\placefigure{fig:energy3}
\begin{figure}[h]

\epsscale{0.82}
\begin{turn}{-90}
\plotone{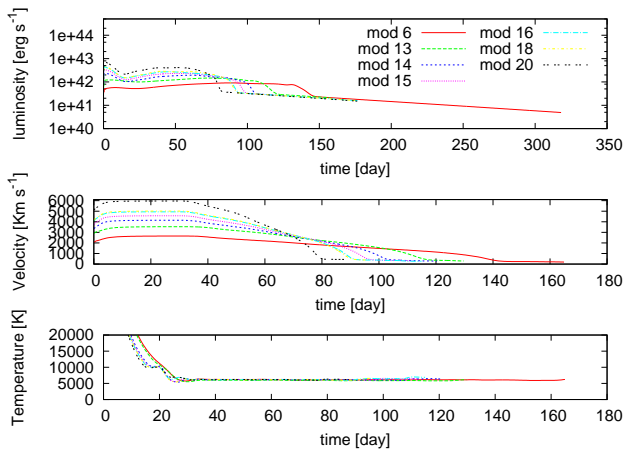}
\end{turn}
\caption{As for Fig.~\ref{fig:mass_env}, but for the total energy. The lines refer to models
(6), (13), (14), (15), (16), (18), and (20) with $E=0.5$, $1$, $1.3$, $1.6$, $1.9$, $2$, and 
$3$ foe, respectively. The models (see Table \ref{tab_models}) have the same initial radius 
($R_0=3\times10^{13}$ cm), envelope mass ($M_{env}=16$\Msun), and amount of \chem{56}{Ni} 
($M_{Ni}=0.07$\Msun).
\label{fig:energy3}}
\end{figure}

\placefigure{fig:ni}
\begin{figure}[h]
\epsscale{0.82}
\begin{turn}{-90}
\plotone{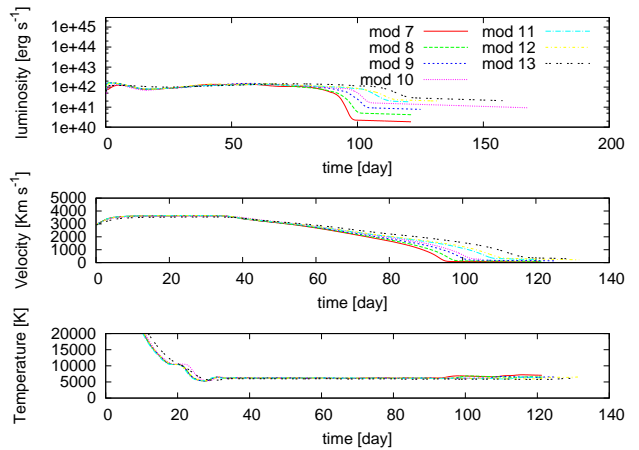}
\end{turn}
\caption{As for Fig.~\ref{fig:mass_env}, but for the \chem{56}{Ni} mass. The lines refer
to models (7), (8), (9), (10), (11), (12), and (13) with $M_{Ni}=0.005$, $0.01$, $0.02$, 
$0.035$, $0.045$, $0.055$, and $0.070$\Msun, respectively. The models (see Table \ref{tab_models}) 
have the same initial radius ($R_0=3\times10^{13}$ cm), envelope mass ($M_{env}=16$\Msun), 
and total energy ($E=1$ foe).
\label{fig:ni}}
\end{figure}

As can be seen from Figure \ref{fig:mass_env}, more massive envelopes lead to a fainter bolometric 
light curve during the diffusive phase, a longer plateau phase, a hotter photospheric temperature 
during the first $\sim$ 20-30 days, a lower photospheric velocity up to $\sim$ 50 days and a slower 
decline at longer times. As known, this behavior is a consequence of the increase in the diffusion 
timescale with increasing envelope mass, that causes a slower leakage of photons and hence a lower 
luminosity, longer plateau and higher temperature. At the same time, increasing the mass and keeping 
the initial kinetic energy constant (which is half of the total energy; see 
Sect.~\ref{sec:simul:models}) results in a smaller photospheric velocity during the first weeks 
and a slower decline later on, caused by the slower motion of the RW in a higher density envelope.

The effects of varying the initial radius are shown in Figure \ref{fig:radius}. Larger values of $R_0$ 
cause a higher luminosity during the diffusive and recombination phases, a longer plateau, and a smaller 
photospheric velocity up to $\sim$ 25 days. The higher luminosity $L$ during the diffusive phase is 
caused by the fact that $L$ is proportional to the initial radius $R_0$ \citep[e.g.][]{arnett96}, as the 
radiative flux $F\propto T^4/(\kappa_{es} \rho_0 R_0) \propto R_0^{-1}$. The lower initial photospheric 
velocity, longer plateau and higher initial temperature are a consequence of the smaller $PdV$ work done 
by bigger expanding envelopes. Less internal energy is initially converted into kinetic energy, the 
expansion is slower and more energy remains available for heating up the envelope.

Figure \ref{fig:energy3} shows the effects of varying the total energy (with ratio between
thermal and kinetic energy equal to 1; see Sect.~\ref{sec:simul:models}). 
Increasing it causes a higher luminosity in the bolometric light curve during the diffusive and recombination 
phases, a shorter plateau, a higher photospheric velocity up to $\sim 60-70$ days and a faster decline at 
longer times. The early behavior can be easily understood as a consequence of the larger internal energy 
dumped in the ejecta, that increases their initial velocity and accelerates all the evolutionary stages.

As for the effects of varying \chem{56}{Ni} (see Figure \ref{fig:ni}), increasing it makes the plateau phase 
longer and causes a slower decline of the bolometric light curve towards the radioactive tail. Furthermore, as 
$M_{Ni}$ increases, also the decline of the photospheric velocity is slower, while the bolometric luminosity 
becomes higher during the radioactive tail. All these effects are related to the increased heating provided by the 
larger amount of \chem{56}{Ni} concentrated in the innermost part of the envelope, that increases the internal 
energy and luminosity especially at late phases and slows down the motion of the RW \citep[e.g.][]{arnett96,hamuy03b,Nadyozhin03}.

\section{Summary and final comments}
\label{sec:summary}

We developed a general-relativistic, radiation hydrodynamics, Lagrangian code tailored to the radiation-hydrodynamical 
modelling of CC-SNe, whose distinctive features are an accurate treatment of radiative transfer coupled to relativistic 
hydrodynamics, a self-consistent treatment of the evolution of the innermost ejecta taking into account the gravitational 
effects of the central compact remnant, and a fully implicit Lagrangian approach to the solution of the coupled non-linear 
finite difference system of equations. The latter property represents the major novelty of this code, that is based on a 
previous version originally developed for studying fall back in the aftermath of a SN explosion \citep[][]{zampieri98,balberg00}.

With this code, a total of 22 models of hydrogen-rich CC-SNe were calculated, enabling us to (a) 
validate the code against observations and similar models reported in the literature, and (b) study 
the role of the ``main'' parameters affecting the post-explosion evolution of the CC-SN events 
(namely the ejected mass, the progenitor radius, the explosion energy, and the amount of 
\chem{56}{Ni}).

The aforementioned grid of models permitted us also to explore possible correlations among 
different quantities that can be measured from the light curve and spectra, and investigate how 
variations of basic parameters combine together to produce them. This is particularly relevant 
in connection with the possibility to calibrate hydrogen-rich CC-SNe to turn them into usable 
distance indicators. Indeed, in recent years growing attention has been devoted to the construction 
of Hubble diagrams using hydrogen-rich CC-SN events \citep[e.g.][]{nugent06,olivares10} in order to 
derive cosmological parameters independently of the usual method based on type Ia SNe
\citep[e.g.][]{riess98,perlmutter99,astier06,wood-vasey07,hicken09,freedman09}. 

Although a complete study using a more extended grid of models calculated from realistic initial 
conditions is presently under investigation, here we mention some preliminary results 
obtained for the sample of 22 models presented in this paper. Indeed, despite the limitations related 
to the use of simplified initial conditions, we are able to reproduce the power-law relation 
between the luminosity and the photospheric velocity (both measured at day 50 from the breakout of the 
shock wave at the stellar surface) found in the observational sample of \citet[][]{HP02}.
The index of the theoretical correlation is equal to $3.13\pm 0.29$, in good agreement within the 
errors with the value $3.03\pm 0.37$ reported by \citet[][]{HP02}. Our models confirm also the 
anti-correlation between the light curve slope at the so-called inflection time $t_i$ (when the 
semi-logarithmic derivative of the luminosity $L$ at the end of the plateau is maximum) and the amounts of \chem{56}{Ni} 
inferred by \citet*[][]{Elmhamdi03} on observational bases. Finally, we found a very promising 
calibration relation between the luminosity $L_*$ at a generic time $t_*$ during the plateau and the characteristic 
time $t_c=t_{0.4}-t_*$, where $t_{0.4}$ is the time when $L_*$ decreases by a factor 2.5. These last two relations 
may represent useful tools for calibrating hydrogen-rich CC-SNe using only photometric data. In particular, for 
type II plateau SNe the latter relation is essentially independent of the explosion epoch as $L_*$ is approximately 
constant during the plateau.

We are working at present to check the validity of these preliminary results against a more extended grid 
of models that is being computed from realistic initial conditions. We plan to check these calibrations also against
observations using well sampled light curves of hydrogen-rich CC-SNe that are being collected within the 
dedicated large program ``Supernova Variety and Nucleosynthesis Yields'' (P.I. S. Benetti) presently running 
at the European Southern Observatory and the Telescopio Nazionale Galileo. The aforementioned grid of models 
will also serve to build an extended database to be compared with observations of single SNe in order to infer 
their physical properties, in analogy to what already being done using models with 
simplified initial conditions (e.g. SN 2007od; see \citealt{inserra11}).

Our medium- and long-term goal is the development of a sort of ``CC-SNe Laboratory'' in which our code 
is interfaced, in input, with other codes dealing with the calculations of the pre-SN evolution 
\citep[see e.g.][and references therein]{LC03} and explosive nucleosynthesis (see e.g. 
\citealt{CL04}, and references therein), and in output with a spectral synthesis code 
\citep[see e.g.][]{mazzali00,ML93}. 
This will allow us to describe the evolution of a CC-SN event in a ``self-consistent'' way from the 
evolutionary stages preceding the main sequence up to the post-explosive phases as a function of 
initial mass, metallicity, stellar rotation, and mass loss history of the CC-SN progenitor.

\appendix
\section{Finite difference equations}
\label{appendix}

We start from Eqs.~[\ref{eq:energy}], [\ref{eq:w0}] and [\ref{eq:w1}], write $r_{,t}=au$, and express 
$b_{,t}/b = - (\rho r^2)_{,t}/\rho r^2$ in terms of the continuity equation (Eqs.~[8] and [4] in 
\citealt{zampieri98}). We then discretize the spatial computational domain (in Lagrangian mass $\mu$) 
dividing it into a grid of $j_{max}$ zones (equal to 110 in our simulations), where a constant fractional 
increment in grid spacing between successive zones $\alpha$ is used. This fractional increment is calculated 
from the following relation:
\begin{equation}
\mu_{max}=\mu_{min} + \sum_{j=j_{min}}^{j_{max}-1} \Delta\mu_{j+1/2} \, ,
\label{finite:mu}
\end{equation}
where $\mu_{min}=\mu_{j_{min}}$ is the inner boundary of the grid in Lagrangian mass, $\mu_{max}=\mu_{j_{max}}$ is 
the outer boundary (fixed during a simulation due to the conservation of the envelope mass), and 
$\Delta\mu_{j+1/2}\equiv \mu_{j+1} - \mu_{j} = \alpha\Delta\mu_{j-1/2}$. At the beginning of the simulation, $\mu_{min}=0$, 
$\mu_{max}=M_{env}$, and the mass contained within the first shell $\Delta\mu_{1/2}$ is fixed by the requirement that 
the radial spacing between the first two shells is $30$\% of the inner radial boundary $r_{in}=r(\mu_{min})$. 
During the evolution, if the inner edge of the innermost zone crosses $r_{in}$, it is removed from the calculation and 
a regridding of all the variables is performed so as to have $j_{max}$ zones at all times. In particular, the new inner 
boundary is set equal to $\mu_{j_{min}+l}$, where $l$ is the number of zones that have crossed $r_{in}$, and the new 
fractional increment $\alpha$ is calculated from Eq.~[\ref{finite:mu}] with the new value of inner boundary. Afterwards, 
all the variables are interpolated on the new grid.

As for the spatial centering of the variables, $B=a_R T^4$ and $w_0$ are evaluated at mid zone ($\mu_{j-1/2}$), while $w_1$ 
at zone boundary ($\mu_j$). Concerning the time centering, all quantities are evaluated at the full time level $t^n$. With 
this centering and following a standard Lagrangian approach for the discretization of the equations and the derivatives, the 
finite difference form of Eqs.~[\ref{eq:energy}], [\ref{eq:w0}] and [\ref{eq:w1}] can be written as:
\begin{eqnarray}
\frac{\epsilon_{j-1/2}^{n+1} - \epsilon_{j-1/2}^n}{\Delta t^{n+1/2}} 
& + & a_{j-1/2}^{n+1/2} (k_P)_{j-1/2}^{n+1/2}
\left[ B_{j-1/2}^{n+1/2} - (w_0)_{j-1/2}^{n+1/2} \right] + \nonumber \\
& & + \frac{p_{j-1/2}^{n+1/2}}{\Delta t^{n+1/2}} \left( \frac{1}{\rho_{j-1/2}^{n+1}} - \frac {1}{\rho_{j-1/2}^n} \right) = 0
\label{finite:energy}\\
\frac{(w_0)_{j-1/2}^{n+1} - (w_0)_{j-1/2}^n}{\Delta t^{n+1/2}} 
& - & a_{j-1/2}^{n+1/2} (k_P)_{j-1/2}^{n+1/2} \rho_{j-1/2}^{n+1/2}
\left[ B_{j-1/2}^{n+1/2} - (w_0)_{j-1/2}^{n+1/2} \right] + \nonumber \\
& & + (w_0)_{j-1/2}^{n+1/2} a_{j-1/2}^{n+1/2} \left[ \left( \frac{4}{3} 
+ f_{j-1/2}^{n+1/2} \right) \frac{1}{(r^2)_{j-1/2}^{n+1/2}} \times \right. 
\nonumber \\
& & \left. \times \frac{ u_j^{n+1/2}(r_j^{n+1/2})^2 - 
u_{j-1}^{n+1/2}(r_{j-1}^{n+1/2})^2 }
{ r_j^{n+1/2} - r_{j-1}^{n+1/2} } - 3 \left(f \frac{u}{r} 
\right)_{j-1/2}^{n+1/2}
\right] + \nonumber \\
& & + \left( \frac{\Gamma}{ar^2} \right)_{j-1/2}^{n+1/2}
\left[ \frac{ (w_1)_j (a_j)^2 (r_j)^2 - (w_1)_{j-1} (a_{j-1})^2 (r_{j-1})^2 }
{ r_j - r_{j-1} } \right]^{n+1/2} + \nonumber \\
& & - \left[ \frac{4\pi r a}{\Gamma} \left( \frac{4}{3} + f \right) w_0 
w_1 \right] _{j-1/2}^{n+1/2} = 0
\label{finite:w0}\\
\frac{(w_1)_{j}^{n+1} - (w_1)_{j}^n}{\Delta t^{n+1/2}} 
& + & \frac{(F_w)^{n+1/2} (w_1)_{j}^{n+1/2}}{\Delta t^{n+1/2}} - 
\frac{(H_w)^{n+1/2}}{\Delta t^{n+1/2}} = 0\,,
\label{finite:w1}
\end{eqnarray}
where
\begin{eqnarray}
B_{j-1/2}^{n+1/2}     & = & \frac{B_{j-1/2}^n + B_{j-1/2}^{n+1}}{2}\\
(w_0)_{j-1/2}^{n+1/2} & = & \frac{(w_0)_{j-1/2}^n + (w_0)_{j-1/2}^{n+1}}{2}\\
F_w                   & = & \Delta t^{n+1/2} a_j^{n+1/2} \left[ (k_R)_j \rho_j + \frac{2}{r_j}
\frac{u_{j+1/2}r_{j+1/2} - u_{j-1/2}r_{j-1/2}}{r_{j+1/2} - r_{j-1/2}} 
\right]^{n+1/2} \\
H_w  & = & \Delta t^{n+1/2} \left\{ \frac{8\pi a_j r_j}{\Gamma_j} (w_1)_j^2 -
a_j \Gamma_j \left[ \frac{ \left( 1/3 + f_{j+1/2} \right) (w_0)_{j+1/2}
- \left( 1/3 + f_{j-1/2} \right) (w_0)_{j-1/2} }{ r_{j+1/2} - r_{j-1/2} }
\right] + \right. \nonumber \\
& & \left. - \Gamma_j \left( \frac{4}{3} + f_j \right) (w_0)_j
\frac{a_{j+1/2} - a_{j-1/2}}{r_{j+1/2} - r_{j-1/2}} +
- 3 \frac{a_j \Gamma_j}{r_j} f_j (w_0)_j \right\}^{n+1/2} \, .
\end{eqnarray}

As $\epsilon$, $p$, $k_R$ and $k_P$ are in general rather complex, non-linear functions of $\rho$ and $B$ 
(see, e.g., Eqs.~[14] and [15] in \citealt{zampieri98}), relations [\ref{finite:energy}]-[\ref{finite:w1}] 
form a highly non-linear system of (3$j_{max}$-3) equations having (3$j_{max}$-3) unknowns, that we solve 
by means of the Newton-Raphson iterative method \citep[see, e.g.,][for details]{NumRec}. Using this approach, 
the finite difference equations are linearized and solved iteratively for the corrections at the new time 
level $t^{n+1}$ starting from an initial guess. The equations for the corrections are:
\begin{displaymath}
\left\{ {\begin{array}{lcl}
\sum_{j_{min}+1}^{j_{max}} \left( \frac{\partial F_{\ref{finite:energy}}(j_{min}+1)}{\partial B_{j-1/2}} \Delta B_{j-1/2} + \frac{\partial F_{\ref{finite:energy}}(j_{min}+1)}{\partial (w_0)_{j-1/2}} \Delta (w_0)_{j-1/2} + \frac{\partial F_{\ref{finite:energy}}(j_{min}+1)}{\partial (w_1)_{j}} \Delta (w_1)_{j} \right) & = & - F_{\ref{finite:energy}}(j_{min}+1) \nonumber \\
\sum_{j_{min}+1}^{j_{max}}\left( \frac{\partial F_{\ref{finite:w0}}(j_{min}+1)}{\partial B_{j-1/2}} \Delta B_{j-1/2} + \frac{\partial F_{\ref{finite:w0}}(j_{min}+1)}{\partial (w_0)_{j-1/2}} \Delta (w_0)_{j-1/2} + \frac{\partial F_{\ref{finite:w0}}(j_{min}+1)}{\partial (w_1)_{j}} \Delta (w_1)_{j} \right) & = & - F_{\ref{finite:w0}}(j_{min}+1) \nonumber \\
\sum_{j_{min}+1}^{j_{max}}\left( \frac{\partial F_{\ref{finite:w1}}(j_{min}+1)}{\partial B_{j-1/2}} \Delta B_{j-1/2} + \frac{\partial F_{\ref{finite:w1}}(j_{min}+1)}{\partial (w_0)_{j-1/2}} \Delta (w_0)_{j-1/2} + \frac{\partial F_{\ref{finite:w1}}(j_{min}+1)}{\partial (w_1)_{j}} \Delta (w_1)_{j} \right) & = & - F_{\ref{finite:w1}}(j_{min}+1) \nonumber \\ \nonumber \\
\nonumber \\
\ldots \ldots \ldots \ldots \ldots \ldots \ldots \ldots \ldots \ldots \ldots \ldots \ldots \ldots \ldots \ldots \ldots \ldots \ldots \ldots \ldots \ldots \ldots \ldots \ldots \ldots \ldots & = & \ldots \ldots \ldots \nonumber \\ \nonumber \\
\sum_{j_{min}+1}^{j_{max}} \left( \frac{\partial F_{\ref{finite:energy}}(j)}{\partial B_{j-1/2}} \Delta B_{j-1/2} + \frac{\partial F_{\ref{finite:energy}}(j)}{\partial (w_0)_{j-1/2}} \Delta (w_0)_{j-1/2} + \frac{\partial F_{\ref{finite:energy}}(j)}{\partial (w_1)_{j}} \Delta (w_1)_{j} \right) & = & - F_{\ref{finite:energy}}(j) \nonumber \\
\sum_{j_{min}+1}^{j_{max}}\left( \frac{\partial F_{\ref{finite:w0}}(j)}{\partial B_{j-1/2}} \Delta B_{j-1/2} + \frac{\partial F_{\ref{finite:w0}}(j)}{\partial (w_0)_{j-1/2}} \Delta (w_0)_{j-1/2} + \frac{\partial F_{\ref{finite:w0}}(j)}{\partial (w_1)_{j}} \Delta (w_1)_{j} \right) & = & - F_{\ref{finite:w0}}(j) \\
\sum_{j_{min}+1}^{j_{max}}\left( \frac{\partial F_{\ref{finite:w1}}(j)}{\partial B_{j-1/2}} \Delta B_{j-1/2} + \frac{\partial F_{\ref{finite:w1}}(j)}{\partial (w_0)_{j-1/2}} \Delta (w_0)_{j-1/2} + \frac{\partial F_{\ref{finite:w1}}(j}{\partial (w_1)_{j}} \Delta (w_1)_{j} \right) & = & - F_{\ref{finite:w1}}(j) \nonumber \\ \nonumber \\
\nonumber \\
\ldots \ldots \ldots \ldots \ldots \ldots \ldots \ldots \ldots \ldots \ldots \ldots \ldots \ldots \ldots \ldots \ldots \ldots \ldots \ldots \ldots \ldots \ldots \ldots \ldots \ldots \ldots & = & \ldots \ldots \ldots \nonumber \\ \nonumber \\
\sum_{j_{min}+1}^{j_{max}} \left( \frac{\partial F_{\ref{finite:energy}}(j_{max})}{\partial B_{j-1/2}} \Delta B_{j-1/2} + \frac{\partial F_{\ref{finite:energy}}(j_{max})}{\partial (w_0)_{j-1/2}} \Delta (w_0)_{j-1/2} + \frac{\partial F_{\ref{finite:energy}}(j_{max})}{\partial (w_1)_{j}} \Delta (w_1)_{j} \right) & = & - F_{\ref{finite:energy}}(j_{max}) \nonumber \\
\sum_{j_{min}+1}^{j_{max}}\left( \frac{\partial F_{\ref{finite:w0}}(j_{max})}{\partial B_{j-1/2}} \Delta B_{j-1/2} + \frac{\partial F_{\ref{finite:w0}}(j_{max})}{\partial (w_0)_{j-1/2}} \Delta (w_0)_{j-1/2} + \frac{\partial F_{\ref{finite:w0}}(j_{max})}{\partial (w_1)_{j}} \Delta (w_1)_{j} \right) & = & - F_{\ref{finite:w0}}(j_{max}) \nonumber \\
\sum_{j_{min}+1}^{j_{max}}\left( \frac{\partial F_{\ref{finite:w1}}(j_{max})}{\partial B_{j-1/2}} \Delta B_{j-1/2} + \frac{\partial F_{\ref{finite:w1}}(j_{max})}{\partial (w_0)_{j-1/2}} \Delta (w_0)_{j-1/2} + \frac{\partial F_{\ref{finite:w1}}(j_{max})}{\partial (w_1)_{j}} \Delta (w_1)_{j} \right) & = & - F_{\ref{finite:w1}}(j_{max}), \nonumber
\end{array} } \right.
\end{displaymath}
where $F_{\ref{finite:energy}}(j)$, $F_{\ref{finite:w0}}(j)$, and $F_{\ref{finite:w1}}(j)$ are the left-hand sides of the 
Eqs.~[\ref{finite:energy}], [\ref{finite:w0}], and [\ref{finite:w1}], respectively and the value of their derivatives is 
calculated numerically. This system of equations can be written in matrix form as:
\begin{equation}
\label{eq:matrix}
\mathbf{A} \cdot \mathbf{x} = \mathbf{b},
\end{equation}
where $\mathbf{A}$ is the matrix of the coefficients (the so-called Jacobian of the system), the raised dot denotes matrix 
multiplication, and $\mathbf{x}$ and $\mathbf{b}$ are the unknown corrections and the known right-hand sides (written as 
column vectors), respectively. They read:
\begin{displaymath}
\mathbf{A} =
\left[ {\begin{array}{c c ccc c c}
\frac{\partial F_{\ref{finite:energy}}(j_{min}+1)}{\partial B_{(j_{min}+1)-1/2}} & \ldots & \frac{\partial F_{\ref{finite:energy}}(j_{min}+1)}{\partial B_{j-1/2}} & \frac{\partial F_{\ref{finite:energy}}(j_{min}+1)}{\partial (w_0)_{j-1/2}} & \frac{\partial F_{\ref{finite:energy}}(j_{min}+1)}{\partial (w_1)_{j}} & \ldots & \frac{\partial F_{\ref{finite:energy}}(j_{min}+1)}{\partial (w_1)_{j_{max}}}\\
\frac{\partial F_{\ref{finite:w0}}(j_{min}+1)}{\partial B_{(j_{min}+1)-1/2}} & \ldots & \frac{\partial F_{\ref{finite:w0}}(j_{min}+1)}{\partial B_{j-1/2}} & \frac{\partial F_{\ref{finite:w0}}(j_{min}+1)}{\partial (w_0)_{j-1/2}} & \frac{\partial F_{\ref{finite:w0}}(j_{min}+1)}{\partial (w_1)_{j}} & \ldots & \frac{\partial F_{\ref{finite:w0}}(j_{min}+1)}{\partial (w_1)_{j_{max}}}\\
\frac{\partial F_{\ref{finite:w1}}(j_{min}+1)}{\partial B_{(j_{min}+1)-1/2}} & \ldots & \frac{\partial F_{\ref{finite:w1}}(j_{min}+1)}{\partial B_{j-1/2}} & \frac{\partial F_{\ref{finite:w1}}(j_{min}+1)}{\partial (w_0)_{j-1/2}} & \frac{\partial F_{\ref{finite:w1}}(j_{min}+1)}{\partial (w_1)_{j}} & \ldots & \frac{\partial F_{\ref{finite:w1}}(j_{min}+1)}{\partial (w_1)_{j_{max}}}\\
\vdots & & \vdots & \vdots & \vdots & & \vdots \\
\vdots & & \vdots & \vdots & \vdots & & \vdots \\
\frac{\partial F_{\ref{finite:energy}}(j)}{\partial B_{(j_{min}+1)-1/2}} & \ldots & \frac{\partial F_{\ref{finite:energy}}(j)}{\partial B_{j-1/2}} & \frac{\partial F_{\ref{finite:energy}}(j)}{\partial (w_0)_{j-1/2}} & \frac{\partial F_{\ref{finite:energy}}(j)}{\partial (w_1)_{j}} & \ldots & \frac{\partial F_{\ref{finite:energy}}(j)}{\partial (w_1)_{j_{max}}}\\
\frac{\partial F_{\ref{finite:w0}}(j)}{\partial B_{(j_{min}+1)-1/2}} & \ldots & \frac{\partial F_{\ref{finite:w0}}(j)}{\partial B_{j-1/2}} & \frac{\partial F_{\ref{finite:w0}}(j)}{\partial (w_0)_{j-1/2}} & \frac{\partial F_{\ref{finite:w0}}(j)}{\partial (w_1)_{j}} & \ldots & \frac{\partial F_{\ref{finite:w0}}(j)}{\partial (w_1)_{j_{max}}}\\
\frac{\partial F_{\ref{finite:w1}}(j)}{\partial B_{(j_{min}+1)-1/2}} & \ldots & \frac{\partial F_{\ref{finite:w1}}(j)}{\partial B_{j-1/2}} & \frac{\partial F_{\ref{finite:w1}}(j)}{\partial (w_0)_{j-1/2}} & \frac{\partial F_{\ref{finite:w1}}(j)}{\partial (w_1)_{j}} & \ldots & \frac{\partial F_{\ref{finite:w1}}(j)}{\partial (w_1)_{j_{max}}}\\
\vdots & & \vdots & \vdots & \vdots & & \vdots \\
\vdots & & \vdots & \vdots & \vdots & & \vdots \\
\frac{\partial F_{\ref{finite:energy}}(j_{max})}{\partial B_{(j_{min}+1)-1/2}} & \ldots & \frac{\partial F_{\ref{finite:energy}}(j_{max})}{\partial B_{j-1/2}} & \frac{\partial F_{\ref{finite:energy}}(j_{max})}{\partial (w_0)_{j-1/2}} & \frac{\partial F_{\ref{finite:energy}}(j_{max})}{\partial (w_1)_{j}} & \ldots & \frac{\partial F_{\ref{finite:energy}}(j_{max})}{\partial (w_1)_{j_{max}}}\\
\frac{\partial F_{\ref{finite:w0}}(j_{max})}{\partial B_{(j_{min}+1)-1/2}} & \ldots & \frac{\partial F_{\ref{finite:w0}}(j_{max})}{\partial B_{j-1/2}} & \frac{\partial F_{\ref{finite:w0}}(j_{max})}{\partial (w_0)_{j-1/2}} & \frac{\partial F_{\ref{finite:w0}}(j_{max})}{\partial (w_1)_{j}} & \ldots & \frac{\partial F_{\ref{finite:w0}}(j_{max})}{\partial (w_1)_{j_{max}}}\\
\frac{\partial F_{\ref{finite:w1}}(j_{max})}{\partial B_{(j_{min}+1)-1/2}} & \ldots & \frac{\partial F_{\ref{finite:w1}}(j_{max})}{\partial B_{j-1/2}} & \frac{\partial F_{\ref{finite:w1}}(j_{max})}{\partial (w_0)_{j-1/2}} & \frac{\partial F_{\ref{finite:w1}}(j_{max})}{\partial (w_1)_{j}} & \ldots & \frac{\partial F_{\ref{finite:w1}}(j_{max})}{\partial (w_1)_{j_{max}}}
\end{array} } \right]
\end{displaymath}

\begin{displaymath}
\mathbf{x} =
\left[ \begin{array}{c}
\displaystyle
\Delta B_{(j_{min}+1)-1/2}\\
\Delta (w_0)_{(j_{min}+1)-1/2}\\
\Delta (w_1)_{(j_{min}+1)} \\
\vdots\\
\Delta B_{j-1/2}\\
\Delta (w_0)_{j-1/2}\\
\Delta (w_1)_{j} \\
\vdots\\
\Delta B_{(j_{max})-1/2}\\
\Delta (w_0)_{(j_{max})-1/2}\\
\Delta (w_1)_{(j_{max})} 
\end{array} \right]
\,\,\,\,\,\,\,\,\,\,\,
\mathbf{b} =
\left[ \begin{array}{c}
\displaystyle
- F_{\ref{finite:energy}}(j_{min}+1)\\
- F_{\ref{finite:w0}}(j_{min}+1)\\
- F_{\ref{finite:w1}}(j_{min}+1)\\
\vdots\\
- F_{\ref{finite:energy}}(j)\\
- F_{\ref{finite:w0}}(j)\\
- F_{\ref{finite:w1}}(j)\\
\vdots\\
- F_{\ref{finite:energy}}(j_{max})\\
- F_{\ref{finite:w0}}(j_{max})\\
- F_{\ref{finite:w1}}(j_{max})
\end{array} \right].
\end{displaymath}

Given the functional dependence of $F_{\ref{finite:energy}}(j)$, $F_{\ref{finite:w0}}(j)$, 
and $F_{\ref{finite:w1}}(j)$ from $B_{j-1/2}$, $(w_0)_{j-1/2}$, and $(w_1)_{j}$, the matrix 
of coefficients $\mathbf{A}$ is a so-called ``band diagonal'' matrix having nonzero elements 
only on the diagonal plus $m_1$ ($=$6 at most) elements immediately to the left of 
(or below) the diagonal and $m_2$($=$4 at most) elements immediately to the right 
(or above it). In other words, the elements of $\mathbf{A}$ are equal to zero if either 
the column index $h$ is greater than the row index $k+m_2$  or the row index $k$ is greater 
than the column index $h+m_1$.

In order to solve the system [\ref{eq:matrix}] in an efficient way, minimizing the CPU 
time and the required storage space, we adopt a matrix inversion package based on a LU 
decomposition, where the matrix $\mathbf{A}$ is stored and manipulated in a so-called 
``compact form'', in which only the non-zero elements are considered \citep[see, e.g.,][for details]{NumRec}.

In addition, in order to increase the numerical accuracy, we use an iterative improvement 
of the solution $\mathbf{x}$. Specifically, we approximate $\mathbf{x}$ with
$\mathbf{x_o}+\delta\mathbf{x_o}$, where $\mathbf{x_o}$ is the solution of the 
system $\mathbf{A} \cdot \mathbf{x} = \mathbf{b}$ and $\delta\mathbf{x_o}$ -- the first-order 
correction to $\mathbf{x_o}$ -- is the solution of the system $\mathbf{A} \cdot \delta\mathbf{x} = (\mathbf{b}-\mathbf{Ax_0})$
\citep[see, e.g.,][for details]{NumRec}.

\acknowledgments
M.L.P. acknowledges financial support from the municipality of Padua through the prize {\it Padova Citt\`a delle Stelle},
and from the Bonino-Pulejo Foundation. We also acknowledge financial support from the PRIN-INAF 2009 ``Supernovae Variety 
and Nucleosynthesis Yields''. We are grateful to the TriGrid VL project and the INAF-Astronomical Observatory of Padua for 
the use of computer facilities. Finally, we thank Stefano Benetti, Enrico Cappellaro, Paolo Mazzali, and Massimo Turatto 
for their useful comments.


\clearpage

\end{document}